\documentclass[sigplan,nonacm]{acmart}
\usepackage{enumitem}
\usepackage{booktabs}
\usepackage{tabularx}
\begin{document}
\title{MXFormer: A Microscaling Floating-Point Charge-Trap Transistor Compute-in-Memory Transformer Accelerator}

\author{George Karfakis}
\affiliation{
  \institution{University of California, Los Angeles (UCLA)}
  \city{Los Angeles}
  \state{CA}
  \country{USA}
}
\email{georgekarfakis@ucla.edu}

\author{Samyak Chakrabarty}
\affiliation{
  \institution{University of California, Los Angeles (UCLA)}
  \city{Los Angeles}
  \state{CA}
  \country{USA}
}
\email{csamyak@g.ucla.edu}

\author{Vinod Kurian Jacob}
\affiliation{
  \institution{University of California, Los Angeles (UCLA)}
  \city{Los Angeles}
  \state{CA}
  \country{USA}
}
\email{jvinod@g.ucla.edu}

\author{Siyun Qiao}
\affiliation{
  \institution{University of California, Los Angeles (UCLA)}
  \city{Los Angeles}
  \state{CA}
  \country{USA}
}
\email{qiaosy@g.ucla.edu}

\author{Subramanian S. Iyer}
\affiliation{
  \institution{University of California, Los Angeles (UCLA)}
  \city{Los Angeles}
  \state{CA}
  \country{USA}
}
\email{s.s.iyer@ucla.edu}

\author{Sudhakar Pamarti}
\affiliation{
  \institution{University of California, Los Angeles (UCLA)}
  \city{Los Angeles}
  \state{CA}
  \country{USA}
}
\email{spamarti@ee.ucla.edu}

\author{Puneet Gupta}
\affiliation{
  \institution{University of California, Los Angeles (UCLA)}
  \city{Los Angeles}
  \state{CA}
  \country{USA}
}
\email{puneetg@ucla.edu}

\renewcommand{\shortauthors}{Karfakis et al.}

\begin{abstract}
The proliferation of Transformer models is often constrained by the significant computational and memory bandwidth demands of deployment. To address this, we present \emph{MXFormer}, a novel, hybrid, weight-stationary Compute-in-Memory (CIM) accelerator that provides high throughput and efficiency for fixed-model inference on large short-sequence Transformers. Our architecture's foundation is the use of ultra-dense Charge-Trap Transistors (CTTs) in Microscaling MXFP4 CIM arrays, uniquely enabling the on-chip storage of up to hundreds of millions of parameters in Fully Weight Stationary (FWS) fashion.

We introduce a statically partitioned design with 12 Transformer blocks connected by a deeply pipelined dataflow. Static-weight layers (MLPs and linear projections) execute on highly parallel analog CTT arrays using an MXFP4-native flow with per-block exponent alignment and a 10-bit SAR ADC. Dynamic computations are handled in fully accurate digital blocks that utilize MXFP-enabled systolic arrays for scaled dot-product attention and vector units for LayerNorm and FlashAttention-style Softmax. 

By eliminating all weight movement, the deeply pipelined \emph{MXFormer} architecture yields very high single-stream throughput and efficiency, processing $58{,}275$ FPS on ViT-L/32 (dual-chip) or $41{,}269$ FPS on ViT-B/16 (single chip). \emph{MXFormer} outperforms comparable state-of-the-art non-FWS digital, hybrid and photonic Transformer accelerators \textbf{$\sim$3.3$\times$--60.5$\times$} in compute density and \textbf{$\sim$1.7$\times$--2.5$\times$} in energy efficiency. Against FWS accelerators, \emph{MXFormer} improves compute density by \textbf{$\sim$20.9$\times$} and resident weight storage density by \textbf{$\sim$2$\times$}, while preserving near-digital accuracy (drop of $\leq1\%$) without \textit{any} model retraining. \end{abstract}

\maketitle

\section{Introduction}
For years, specialized model families with distinct computation patterns shaped accelerator design: sequential RNNs for NLP \cite{original_rnn} and spatially parallel CNNs for vision \cite{original_cnn}. The Transformer \cite{attn_all_you_need} has largely superseded both \cite{transformer_survey_universal}. Introduced for NLP \cite{transformernlp}, it now dominates vision as well, underpinning models such as the Vision Transformer (ViT) \cite{original_vit_paper} and CLIP \cite{openaiCLIP}. Unlike autoregressive large language models (LLMs) that process very long, variable-length sequences, these models use fixed, relatively short token sequences, forming a distinct workload class.

A parallel trend in hardware design is the move towards more efficient numerical formats that balance the high cost of floating-point with the limited dynamic range of integer data types. Microscaling (MX) formats, such as MXFP4 \cite{microsoft_microscaling,ocp_microscaling}, have emerged as a powerful compromise. By combining near-half-precision floating-point range with very high hardware efficiency, these formats have gained significant industry adoption (NVIDIA Blackwell \cite{BlackwellMXFP}, AMD CDNA 4 \cite{cdna4MXFP}, Microsoft Maia 100 \cite{maia100MXFP}) and have been demonstrated to maintain very high accuracy, especially for ViTs \cite{tetrajetmxfp4qat}. 

These trends motivate a core question: given the prevalence of short-sequence Transformers and MXFP4’s efficacy, can a \textit{fully weight-stationary (FWS)} \cite{IBM_nvm_talk} MXFP4-centric architecture be developed to deliver exceptional throughput? A FWS architecture keeps all parameters on-die, removing the dominant off-chip weight traffic \cite{dnn_accel_survey,dnn_mem_survey} and need for external weight memory. We posit that high-density Non-Volatile-Memory (NVM) for weight storage is crucial to achieving this.

Using NVM only as storage still incurs a significant energy/latency penalty to shuttle weights to digital compute \cite{eyeriss,diannao}. We therefore pair NVM storage with \textbf{compute} via \emph{Compute-In-Memory} (CIM): executing the static layers’ matrix-vector operations in-array minimizes on-chip movement, boosting throughput and energy efficiency. FWS CIM also yields fixed, predictable latency, which is important for safety-critical ViT deployments (e.g., automotive) \cite{automotive_transformers}.

However, incorporating NVM-based CIM with a static FWS design that holds all weights on-die is difficult. While ReRAM/PCM/FeRAM are dense, they typically cannot fit large Transformer models within a practical die area \cite{IBM_nvm_talk}, requiring designs with Partially Weight Stationary (PWS) execution and off-chip memory. 
To address this density challenge, we utilize Charge-Trap Transistors (CTTs) \cite{CTT_multibit_2019,ctt_14nm,ctt_cim_iedm, og_ctt_mnist, og_ctt_paper_ibm}, conventionally fabricated 1T non-volatile cells that store up to six bits \cite{ctt_6bit} via charge trapping in a high-$\kappa$ gate oxide. We contend that their exceptional density, accuracy, and low read latency are key technological enablers for a truly weight-stationary CIM architecture for large models.

CTTs are inherently low-noise, highly tunable devices. By combining their high density with a high-bitwidth Analog-to-Digital Converter (ADC), our architecture achieves near-digital accuracy during execution. This enables the deployment of models \emph{without} any of the specialized fine-tuning or retraining steps required by many analog accelerator designs \cite{hyflexpim, photonic_accel_transformer, ucsd_hybrid_attn, ibm_pcm_2023, IBM_nvm_talk, ibm_pcm_pws, imc_qat_noise_1, imc_qat_noise_2}.

While CTT arrays excel for operations with static weights, Transformer attention is a dynamic operation with input-dependent weights, and thus infeasible in weight-stationary NVM \cite{memory_is_all_you_need_survey}. Hence, we adopt a hybrid system design comprising analog CTT for static layers and conventional MXFP4 (with BF16 accumulation) systolic arrays for the attention path.

This paper makes the following contributions:
\begin{itemize}
\item We present a fully weight-stationary, deeply pipelined Transformer accelerator that maps static operations to dense analog CTT CIM macros and dynamic operations to digital systolic arrays and vector units.
\item We develop an MXFP4-native analog CTT CIM macro with per-block exponent alignment across long analog accumulations and a 10-bit SAR ADC, delivering near-digital accuracy \textit{without} Quantization-Aware-Training (QAT).
\item We provide a comprehensive system evaluation and characterization on a 22nm node, demonstrating its high throughput and energy efficiency on multiple short-sequence Transformer models.
\end{itemize}

The remainder of this paper is organized as follows. Section 2 provides a background on the relevant technologies and workloads. Section 3 details the operation and design of the analog components of the system. Section 4 presents the full system architecture. Section 5 presents our evaluation methodology and results. Section 6 compares to prior work, and Section 7 concludes the paper.

\section{Background and Motivation}

\subsection{Transformer}
\subsubsection{Transformer Architecture}
The Transformer is a homogeneous neural network architecture that processes sequences through a series of identical blocks \cite{attn_all_you_need}. Its core, self-attention, assigns data-dependent weights to all inputs (tokens) when forming each token's representation. It captures complex long-range dependencies and underpins state-of-the-art models across domains \cite{original_vit_paper}.

\begin{figure}[!htbp]
    \centering
    \includegraphics[width=0.43\textwidth]{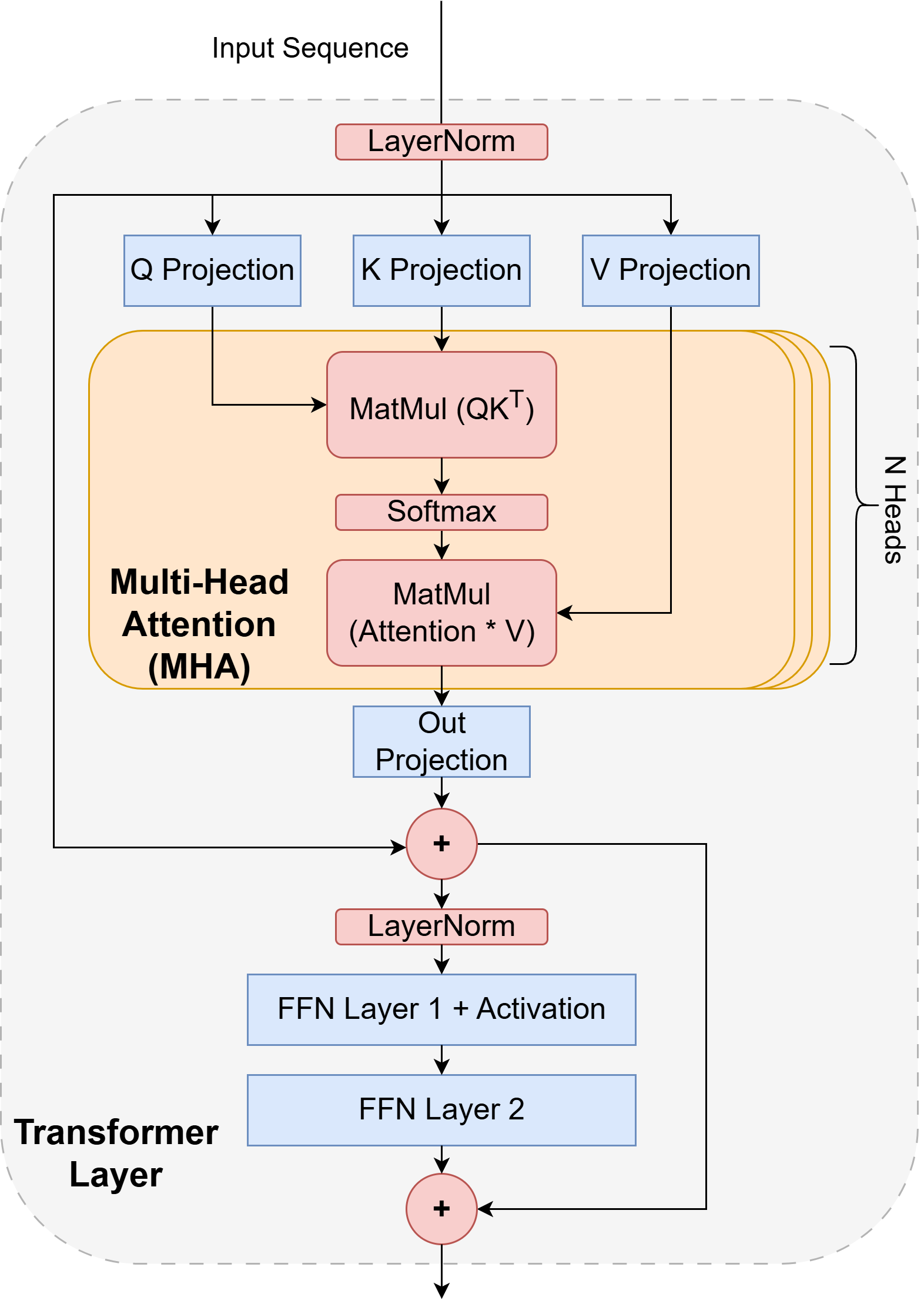}
    \caption{Transformer Layer. Blue components represent linear layers with analog NVM-mappable static weights. Red components utilize dynamic weights and require digital compute units. Layers are connected sequentially to form a full Transformer.}
    \label{fig:transformer}
\end{figure}

As shown in Fig.~\ref{fig:transformer}, each Transformer block is made of two main stages. In the first stage, the input sequence is linearly projected into Query ($\mathbf{Q}$), Key ($\mathbf{K}$), and Value ($\mathbf{V}$) matrices and then partitioned into heads. Then, during the \emph{Multi-Head Attention} (MHA) operation, each head generates attention weights from the scaled dot product $\mathbf{Q}\mathbf{K}^{\top}$, producing a weighted sum of $\mathbf{V}$. The stage finishes with a final output projection. Critically, both operands in this MHA step are activation-dependent, which means that they are dynamic and unknown a priori. The second stage is a \emph{Feed-Forward Network (FFN)}, which is a two-layer MLP with a nonlinearity. Unlike MHA, the FFN and all projection matrices use static, pre-trained weights during inference. To stabilize the network, residual connections and LayerNorm are applied.

The two stages scale differently with sequence length $N$. FFN and linear projections apply the same operation per token, resulting in a linear cost of computation of $O(N)$. MHA relates every token with every other token through dot-product attention, leading to a quadratic cost of $O(N^2)$ \cite{attn_all_you_need}. This linear-quadratic contrast creates a \textbf{sharp divide} in hardware requirements across applications as $N$ varies.

\subsubsection{Transformer Workloads and Target Domain} \phantom{.}\newline
This scaling difference yields two workload classes:

One class is defined by \textbf{LLMs}, which are autoregressive and decoder-only, typically operating on very long and variable-length sequences. In this domain, the $O(N^2)$ attention is the dominating factor, and efficient inference relies on large KV caches \cite{kv_cache_motivator}.

In contrast, the second class consists of \textbf{ViTs and other related models}, which are non-autoregressive and tend to handle short, fixed-length sequences. For example, the encoder-only ViT  \cite{original_vit_paper} converts an image into patches, resulting in short sequence lengths (e.g., $ N{<}200$). CLIP \cite{openaiCLIP} and similar encoder-decoder models share this short-sequence regime. Despite the recent focus on LLMs, these short-sequence models remain widely deployed for many critical applications, such as classification, image generation, visual search, and automotive perception \cite{transformer_survey_universal,automotive_transformers,efficient_transformer_survey, clip_gen, hierarchial_image_gen_CLIP_openai, walmart_clip}.

\begin{figure}[!htbp]
    \centering
\includegraphics[width=0.47\textwidth]{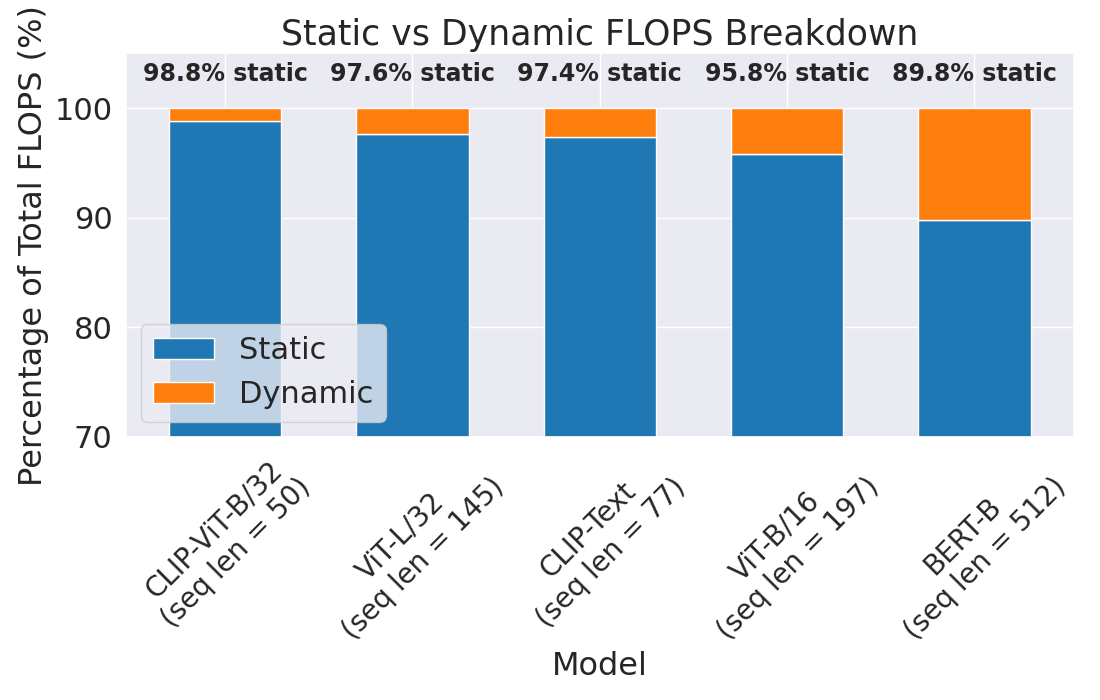}
    \caption{Comparison of Static (mappable to NVM CIM) vs Dynamic (requires digital compute) for various short-sequence length models ($y$-axis starts at 70\%). The sequence length considered is the maximum supported by each model.}
    \label{fig:static_vs_dynamic}
\end{figure}

We focus exclusively on this second class of models. As Fig.~\ref{fig:static_vs_dynamic} shows, static FFN and projection FLOPs dominate short-sequence workloads, making them ideal for a hybrid analog-digital design. While compact digital co-processors handle the smaller dynamic attention path, analog arrays can accelerate the expensive static linear layers. These models also omit the need for a KV cache, enabling fully on-chip inference without external memory.

\subsection{Fully Weight Stationary (FWS) execution}
A Fully Weight Stationary (FWS) accelerator stores the entire model on-die and executes all static operations in place, eliminating on- and off-chip weight traffic entirely. We quantify the resulting reduction in off-chip I/O.

Conventional accelerators/GPUs must re-read weights (amortized by batch) and shuttle activations across layers. The optimal batch is capped by the on-chip activation working set. On an NVIDIA A100 with \textbf{30\,MB} of L2 reserved for persistent data~\cite{cuda_a100_l2_30mb}, Table~\ref{tab:fws_penalty} reports, for each model, (i) the maximum batch whose \emph{activation} set fits in 30\,MB and (ii) the minimum per-item I/O penalty at that batch and at batch size 1.

\begin{table}[!htbp]
\centering
\small
\caption{I/O penalty (additional bytes transferred) \textit{vs.}\ an FWS accelerator. Measured at the max batch that fits in 30\,MB persistent L2 on A100, and at batch size 1.}
\label{tab:fws_penalty}
\begin{tabular}{lcc}
\toprule
Model & Penalty [max batch] & Penalty [batch size = 1] \\
\midrule
BERT-Base & 1.93$\times$ [B=150] & 140$\times$  \\
BERT-Large & 3.86$\times$ [B=112] & 320$\times$  \\
ViT-Base/16 & 1.73$\times$ [B=391] & 285$\times$  \\
ViT-Base/32 & 1.73$\times$ [B=1542] & 1120$\times$ \\
ViT-Large/32 & 3.59$\times$ [B=398] & 1029$\times$ \\
\bottomrule
\end{tabular}
\end{table}

FWS removes this penalty, leaving only a predictable activation stream and mapping naturally to a fixed-latency pipeline with very high single-stream throughput \cite{IBM_nvm_talk}.

\subsection{Microscaling Data Formats}
Microscaling (MX) is an open data format designed for accurate yet efficient execution of AI workloads \cite{ocp_microscaling, microsoft_microscaling}. It represents a length-$k$ vector $V$ as low-precision \textbf{private elements} $P_i$ and a shared floating-point \textbf{scale} $X$ ($V_i = P_i \cdot X$). This preserves high block dynamic range via $X$ while keeping most arithmetic low-bit.

We use the \textbf{MXFP4} format, which uses an 8-bit exponent-only shared scale (E8M0) for every block of 32 4-bit E2M1 private elements (FP4). To enable floating-point semantics on integer analog blocks, we losslessly encode FP4 values as INT5 numbers via an affine map into $[0,24]$ (weights) or $[-12,12]$ (activations) at the boundary of these blocks.

In \emph{MXFormer}, all weights and most activations are stored in MXFP4. Consistent with Microsoft’s implementation \cite{microsoft_microscaling_lib}, numerically sensitive non-linear kernels, such as Softmax, GELU, LayerNorm, and attention accumulations are computed in BF16 \cite{google_bf16}. They are re-quantized back to MXFP4 before being stored in a buffer or entering a linear or attention layer. Conversion details appear in Appendix~\ref{app:conv}.

\subsection{Charge Trap Transistor}
\subsubsection{Device Characteristics}
A Charge-Trap Transistor (CTT) is a non-volatile memory element implemented as a standard high-$\kappa$ nFET in a conventional SOI process. Joule self-heating under moderate bias (e.g., $V_{DS}{=}1.2\mathrm{V}$) induces charge trapping in the gate dielectric, shifting the threshold voltage $V_{th}$ in a programmable, retained manner \cite{CTT_multibit_2019, ctt_cim_iedm}. We encode weights in $V_{th}$ and read them by applying fixed gate/drain biases in subthreshold, so that the drain current represents the stored value. Fine-grained charge programming enables up to six bits per 1T cell \cite{ctt_6bit}. We only utilize five bits per device for this work.

\subsubsection{CTT for Analog Compute-in-Memory}
The MXFP4 format is represented as an unsigned INT5 with a fixed bias. Each CTT cell stores one 5-bit weight. Under binary-decomposed inputs, a logic '1' on the word-line enables the device, sourcing a current proportional to its stored value, implementing an in-situ 5b$\times$1b multiply.

In a crossbar, word-lines drive the input vector to CTT gates while drain currents sum on the shared bit-line, yielding the inner-product. A per-column high-precision ADC digitizes the accumulated current, inherently quantizing the result. Thus each column computes the dot product of its resident weight vector with the streamed input, and columns operate in parallel to deliver thousands of MACs per cycle \emph{in place}, with no weight movement.

\subsubsection{Comparison with Other Analog NVM Technologies}
As shown in Table~\ref{tab:nvme_comp}, CTT as a memory offers the highest storage density among the options considered with a competitive read latency ($\sim$7.5ns). Relative to ReRAM, PCM, FeRAM, CTT can deliver $\geq$1.5$\times$ higher density while maintaining low read energy and access time.

A further advantage is closed-loop programmability. Because a CTT cell is highly tunable and can be read during programming, we use a write-verify-write \cite{ctt_cim_iedm} procedure to dial each cell to target levels. This compensates  for mismatch errors in the read path (e.g., from ADCs and current mirrors), yielding high effective linearity and accuracy. Combined with the CTT’s large signal margins and low read noise, this calibration can create a very accurate system.

\begin{table}[!htbp]
\centering
\small
\caption{NVM comparison. Cell size and Read latency by IRDS for non-CTT NVMs. Max bits-per-cell reported in literature (not same as IRDS figures). $^\dagger$CTT size from tested device in GF22FDX SOI. Latency from post-layout extraction.}
\label{tab:nvme_comp}
\begin{tabular}{@{}lcccc@{}}
\toprule
\textbf{Technology} & \shortstack{Cell Size\\(F$^2$)\\\cite{irds2023}} & \shortstack{Read Latency\\(ns)\\\cite{irds2023}} & \shortstack{Max Bits\\per Cell} & \shortstack{Specialized\\Fabrication} \\
\midrule
NOR Flash & 10 & 50 & 3\cite{nor3} & \textcolor{red}{Yes} \\
ReRAM     & 4--50 & 10--20 & 4\cite{rram4} & \textcolor{red}{Yes} \\
FeRAM     & 10--32 & 20--50 & 3\cite{feram3} & \textcolor{red}{Yes} \\
PCM       & 4--50 & 5--20 & 4\cite{pcm4} & \textcolor{red}{Yes} \\
\textbf{CTT}$^\dagger$ &  5 & 7.5 & 6\cite{ctt_14nm} & \textbf{\textcolor{green}{No}} \\
\bottomrule
\end{tabular}
\end{table}

\section{MXFormer CTT-CIM Macro Architecture}
\subsection{Macro Topology and Organization}

\begin{figure*}[h]
\centering
\includegraphics[width=0.8\textwidth]{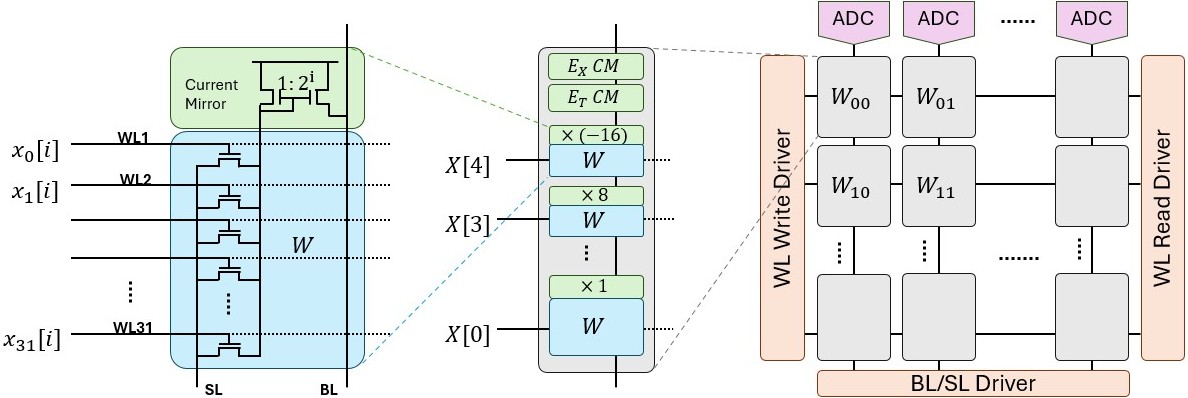}
\caption{Left-to-right: a) Weight block, b) MXFP Block, and c) Macro architecture of the proposed CTT CIM.}
\label{fig:macro_topology}
\end{figure*}

Our macro is a current-summing analog CIM. The bit-cell is an unsigned multi-bit NVM cell (CTT). Inputs are applied via binary decomposition and outputs are digitized by a SAR ADC.

Fig.~\ref{fig:macro_topology} shows the macro and its building blocks. Each output channel (terminating at an ADC) connects to multiple MXFP blocks (Fig.~\ref{fig:macro_topology}b). An MXFP block holds a 32-element INT5 weight vector $W$, consumes a 32-element INT5 input vector $X$, and produces the dot product $Y$. Internally, it instantiates five identical 32-tall weight blocks (Fig. \ref{fig:macro_topology}a), one per input bit-plane $X[4],\dots,X[0]$, and includes two current mirror stacks to adjust input and weight exponents. These mirrors are calibrated using a combination of standard and CTT-assisted procedures.

Let $X=\{x_{31},\dots,x_0\}$ and $W=\{w_{31},\dots,w_0\}$, with binary-decomposed inputs $x_i=\sum_{j=0}^{4}2^j x_i[j],\;x_i[j]\in\{0,1\}$. Then
\begin{align}
Y &= X\cdot W = \sum_{i=0}^{31} x_i\,w_i = \sum_{j=0}^{4} 2^j \!\left(\sum_{i=0}^{31} x_i[j]\cdot w_i\right)
= \sum_{j=0}^{4} 2^j T_j,
\end{align}
where $T_j \triangleq \sum_{i=0}^{31} x_i[j]\cdot w_i$. The weight block (Fig.~\ref{fig:macro_topology}a) comprises 32 five-bit CTT cells tied to 32 word-lines and a shared bit-line; it receives the $j$-th input bit-plane and computes $T_j$. The MXFP block then forms $Y$ from $\{T_j\}$. In practice, we multiplex multiple CTTs connected on different bit-lines to amortize the biasing circuits and ADC, improving weight-storage density. However, only one bit-line is logically active at a time. This lowers analog throughput by the degree of multiplexing.

The macro periphery includes WL/BL drivers for level shifting and biasing, programming drivers for high-voltage writes, and flip-flops to store mirror calibration data.

\begin{table}[!htbp]
\centering
\small
\captionof{table}{Performance summary for macros used. Assumes 1 pass per array. A 2-Pass approach (see Sec. \ref{sec:2pass}) halves all metrics except area.}
\label{tab:performance}
\begin{tabular}{@{}lcc@{}}
\toprule
Technology          & \multicolumn{2}{c}{22\,nm FDSOI} \\
Multiplexing Degree & \multicolumn{2}{c}{10} \\
I/W/O bits          & \multicolumn{2}{c}{5/5/10} \\
Clock Speed (MHz)         &
\multicolumn{2}{c}{169.0} \\
\midrule
\textbf{Array Size} & 768 $\times$ 768 & 1024 $\times$ 1024 \\
Macro Area (mm$^2$) & 1.78 & 2.97 \\
TOPS                & 20.02 & 35.72 \\
TOPS/W              & 58.83 & 75.72 \\
TOPS/mm$^2$         & 11.26 & 12.02 \\
\bottomrule
\end{tabular}
\end{table}

\subsection{Bias Handling \& MXFP Exponent Alignment}

\noindent\textbf{Representation and bias.}
Weights are unsigned INT5 values with a fixed bias $w_b$ that must be removed from the product:
\begin{equation}
    Y \;=\; \sum_i x_i\,(w_i + w_b)
      \;=\; \sum_i x_i\,w_i \;+\; w_b \sum_i x_i.
    \label{eq:bias_adjustment}
\end{equation}
The second term is generated via an identical ``bias'' column with nominal weights and subtracted \emph{per output channel} (not per block) since each column’s effective $w_b'$ differs due to block-specific scaling.

We stream the signed inputs as INT5 two’s-complement bit-planes. Because inputs are already binary-decomposed, flipping the sign of the top mirror (via an NMOS stage) implements the two’s-complement contribution, yielding the correct dot product.

\begin{figure}[!htbp]
\centering
\includegraphics[width=0.3\textwidth]{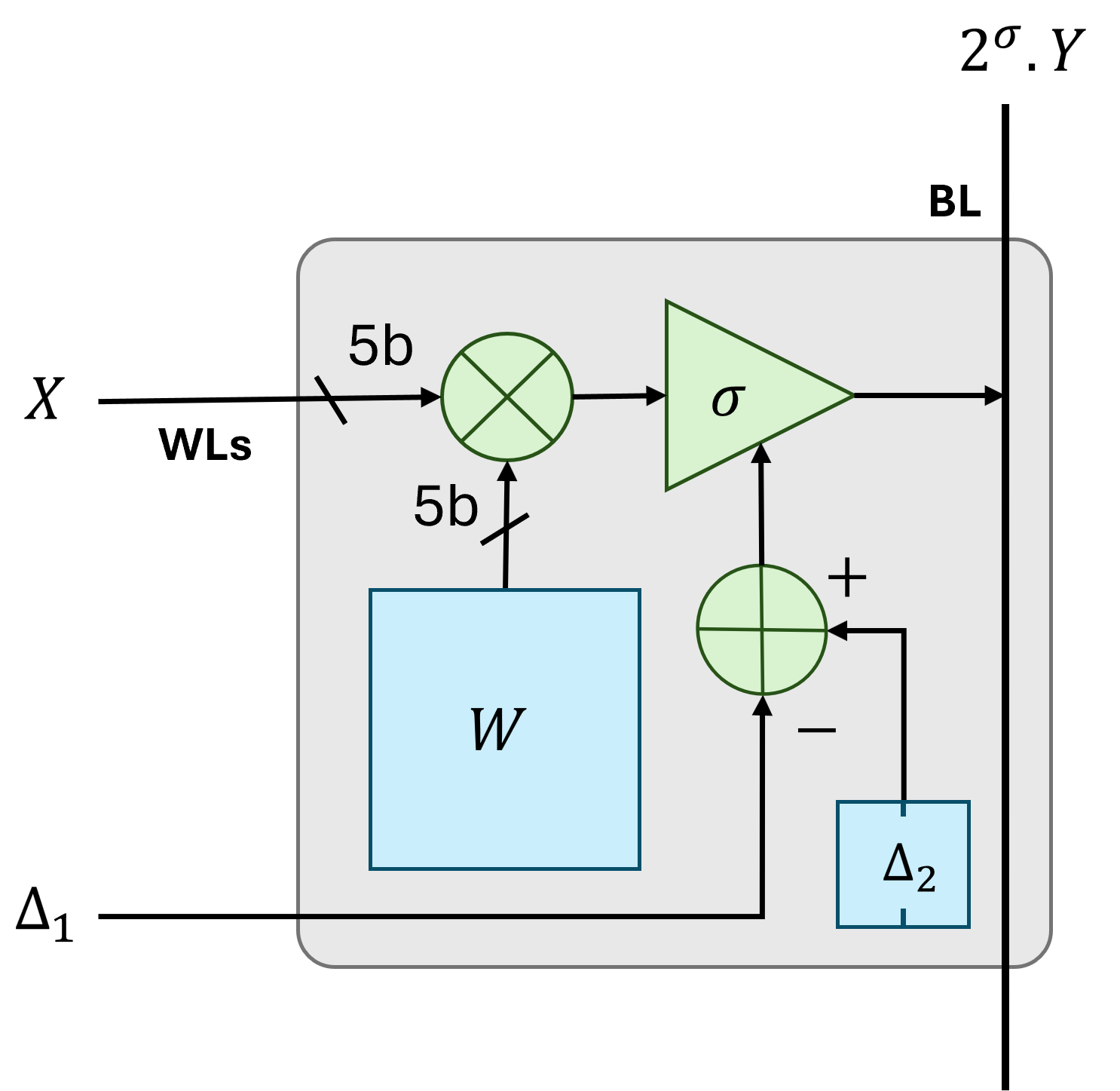}
\caption{Logical schematic of the MXFP block.}
\label{fig:exponent_compensation}
\end{figure}

\noindent\textbf{Block-level exponent alignment.} Intuitively, different MXFP blocks naturally produce partial sums at different shared exponent scales. To sum the contributions of different blocks together on the bit-line, we need to align them to a common target exponent $E_N$. Let $E_W$ be the (static) weight block exponent and $E_X$ the (dynamic) input block exponent. Define $E_T \triangleq E_N - E_W$ (static). The runtime shift is
\begin{equation}
\sigma \;=\; E_N - E_X - E_W \;=\; E_T - E_X,\quad \text{with }\;\; -CM \le \sigma \le 0
\end{equation}

To limit the number of costly, fully-accurate 10-bit current mirror calibration settings, we cap the runtime mirror range to $CM$, and bound $E_T$ to $[E_{\min},\,E_{\min}{+}k]$, which clips $E_X$ to $[E_{\min} -CM,E_{\min} + k]$. As shown in Fig.~\ref{fig:exponent_compensation}, these limited settings are encoded as two scaling factors, $\Delta_1 \in [0,\,CM{+}k]$ (streamed with input) and $\Delta_2 \in [0,\,k]$ (static). Limiting $CM$ (mirror range) and $k$ ($E_T$ window) can introduce clipping when ideal exponents fall outside the allowed ranges. We set $k{=}7$ so $E_T$ remains in-window for the models tested. The choice of $E_N$ and the impact of $CM$ (referred to as the CM Correction bits) are analyzed next.

\subsubsection{Exponent Target Selection Strategies}
\label{sec:2pass}
To choose $E_N$, we use an \textbf{activation-aware offline} ``Row Hist'' strategy. In a one-time calibration, a representative dataset (we used 5 batches) is run to gather the distribution of output activation exponents for every block in each Linear layer. We prioritize avoiding overflows over underflows, as the former diminishes high-magnitude (and thus more critical) activations. Accordingly, we select a \emph{per-layer} $E_N$ that statistically centers the highest expected results within the current mirror budget to eliminate overflows.

\begin{figure}[!htbp]
\centering
\includegraphics[width=0.47\textwidth]{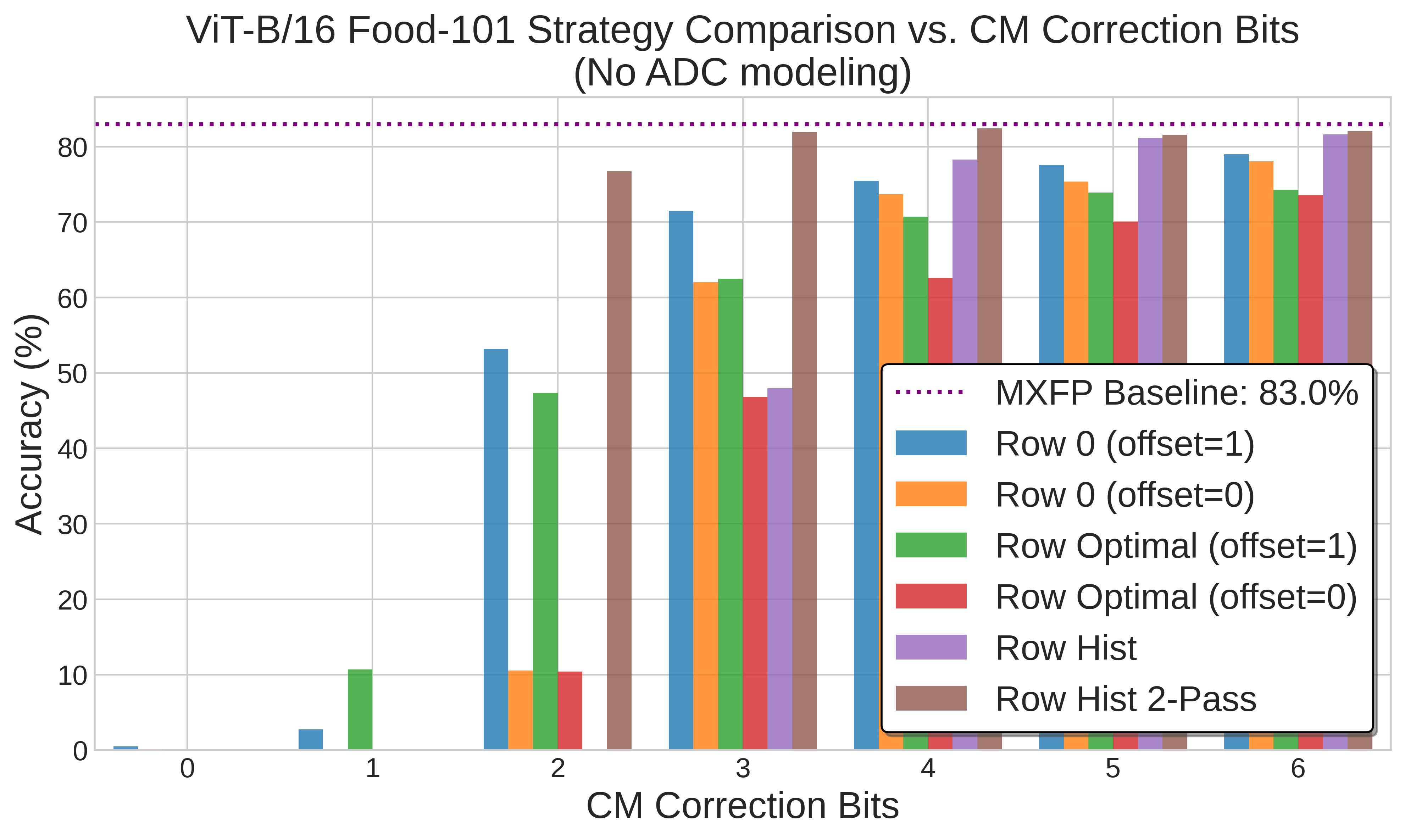}
\caption{A comparison of online and offline exponent target selection strategies. As shown, ``Row Hist 2-Pass'' is effectively identical to ``Row Hist'' at half the CM Correction Bits. The ADC is not modeled.}
\label{fig:strategy_comparison}
\end{figure}

As shown in Fig.~\ref{fig:strategy_comparison}, we also evaluated \textbf{online} alternatives, but they underperformed and were harder to realize in hardware. Using the first block-row’s result exponent for all rows (``Row 0'') or the per-column median shared exponent (``Row Optimal'') reduced accuracy significantly. Adding a constant offset to bias toward larger activations did not close the gap to near-digital results.

\begin{figure}[!htbp]
  \centering
  \begin{minipage}[c]{0.49\columnwidth}
    \centering
    \includegraphics[width=\linewidth]{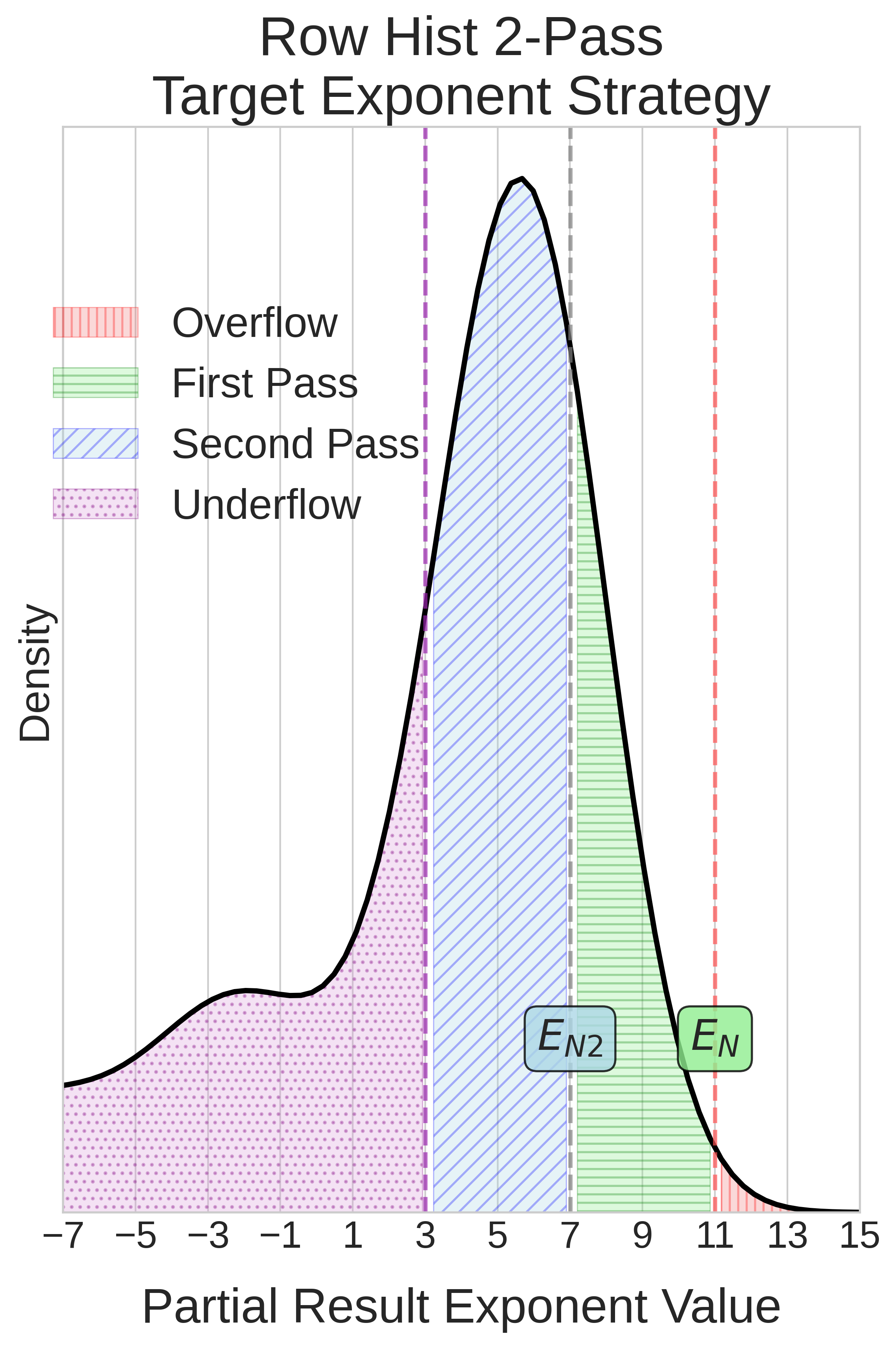}
  \end{minipage}\hfill
  \begin{minipage}[c]{0.49\columnwidth}
    \centering
    \includegraphics[width=\linewidth]{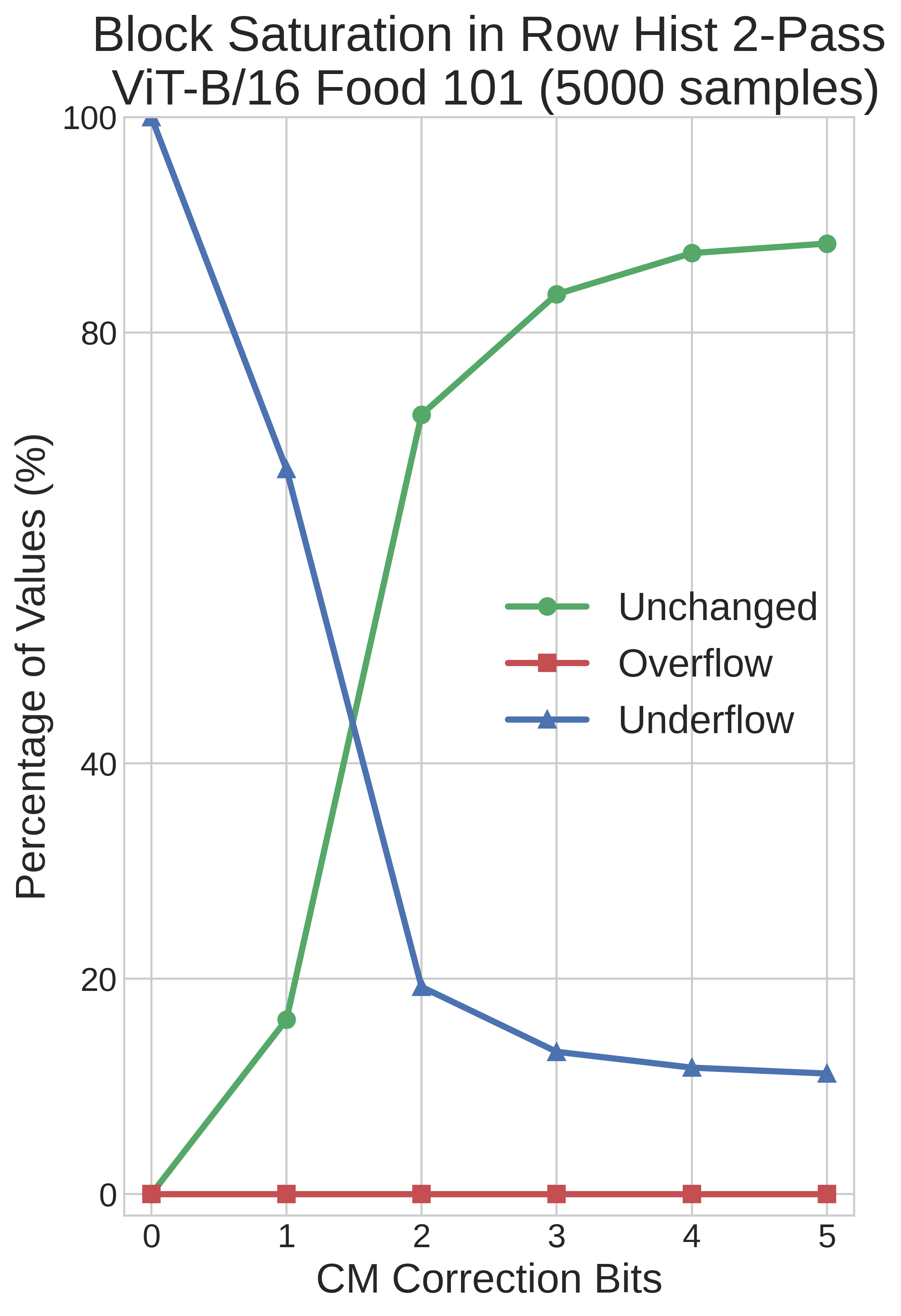}
  \end{minipage}
  \caption{(Left): Illustration of exponent target selection for a 2-pass strategy. As shown, we aim to minimize the overflow region by aligning $E_N$ to the maximum value the shared exponent output assumed during representative benchmarking for each layer. $E_{N2}$ represents the target exponent for the second pass. (Right): Analysis of block shared exponent saturation for the offline Row Hist 2-Pass strategy. For CM correction bits $\geq 3$, the vast majority of activations are preserved.}
  \label{fig:2pass_overflow_combined}
\end{figure}

With ``Row Hist'' at least 5 CM correction bits are needed for near-digital accuracy, which is impractical for dense current mirrors (see Section \ref{sec:cm}). We therefore use a \textbf{2-pass} scheme that extends the effective range to $2\times CM$. In Pass 1, we compute with target $E_N$ and tag blocks that underflow. In Pass 2, only tagged blocks are recomputed with the target $E_{N2} = E_N-CM$, and their results are merged. This is shown in Fig.~\ref{fig:2pass_overflow_combined}. As shown in Fig.~\ref{fig:strategy_comparison} and Table~\ref{tab:acc_res}, this achieves $\leq1\%$ accuracy drop with $CM\!\ge\!3$ using feasible mirrors, at the cost of 50\% lower analog throughput and efficiency.

Fig.~\ref{fig:2pass_overflow_combined} further quantifies how CM correction bits affect saturation across blocks. Our target-exponent strategy eliminates overflow events entirely. For $CM\!\ge\!3$, since at most $16\%$ of blocks underflow to zero, $84\%$ of the higher-magnitude activations are preserved, which explains the minimal accuracy loss.


\subsection{Extracting data from the macro}
The CTT-CIM macro emits 10-bit signed values, which on-chip logic rescales/quantizes to MXFP4. The $\mathbf{Q}$, $\mathbf{K}$, and FFN outputs are quantized row-major in parallel. The $\mathbf{V}$ projection is handled differently. To avoid a large $hidden\_dim\times32$ buffer, we store it as raw INT10 in SRAM and defer its column-wise MXFP4 quantization to the point of use (see Section~\ref{sec:systolic_array}).

\subsection{Analog Design Space Exploration}
We next briefly discuss the area and performance impacts of the analog design decisions we have made, as well as quantify the final model accuracy we achieve. The details behind the accuracy evaluation framework we utilized are explained in Section \ref{sec:software}.

\subsubsection{Current Mirrors}
\label{sec:cm}
Current mirror accuracy improves with device size. Furthermore, each additional CM correction bit approximately doubles the mirror area, creating an exponential area-range trade-off. Extra bits also add calibration states (and flip-flops), limiting how many settings are practical. We found that a correction budget of 3 bits is the maximum we can feasibly support while ensuring lossless current mirror accuracy at a reasonable area.

\begin{figure}[!htbp]
    \centering
    \includegraphics[width=0.48\textwidth]{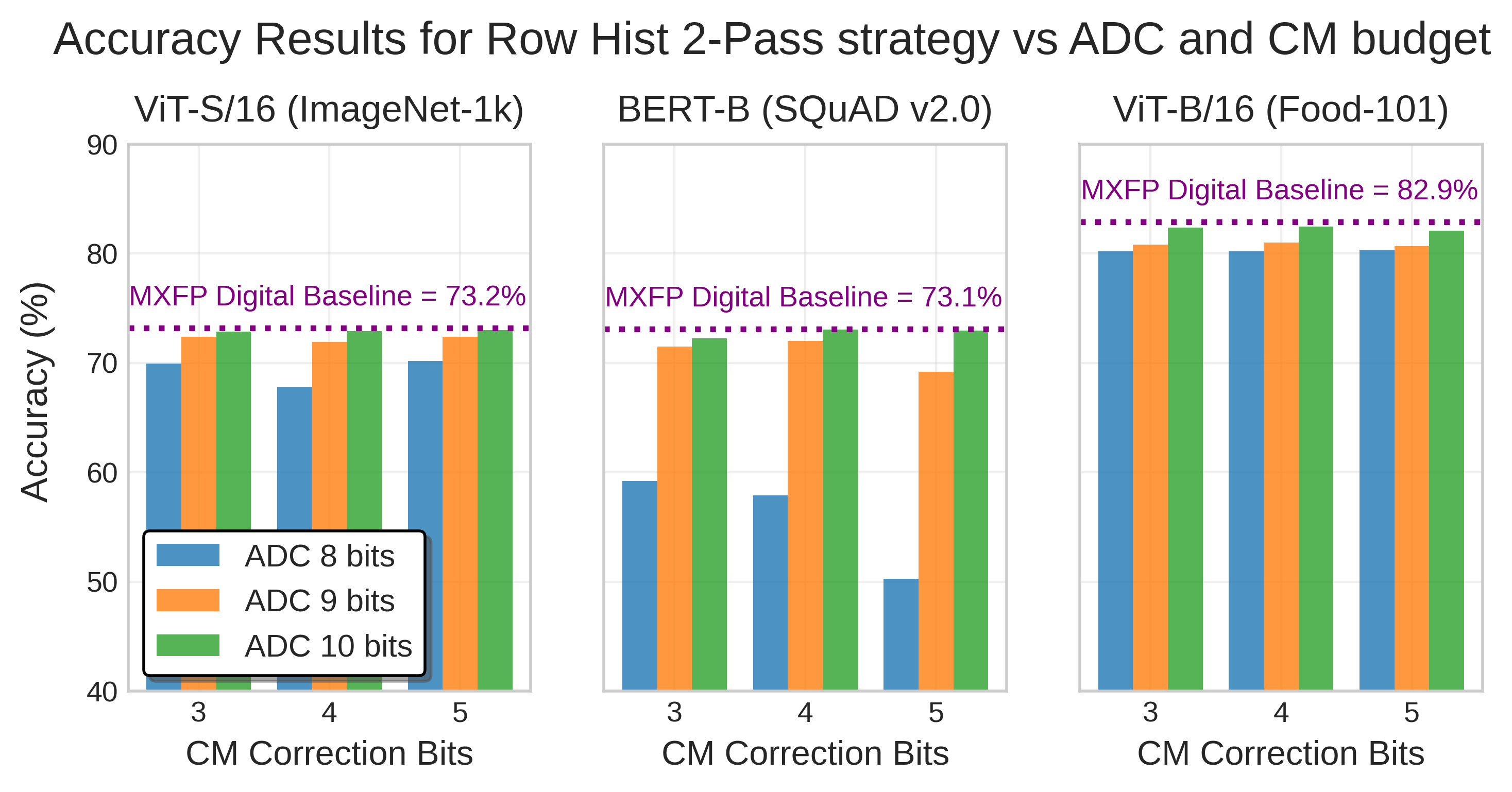}
    \caption{Impact of ADC resolution and overall analog flow on model accuracy. The dotted line is the all-digital MXFP4 baseline. 10 bits are needed for near-digital accuracy in all cases. Y-axis begins at 40\%.}
    \label{fig:adc_accuracy_vitb}
\end{figure}

\subsubsection{Analog to Digital Converter (ADC)}
Under the Row-Hist 2-pass scheme, accuracy improves with ADC resolution and effectively saturates at 10 bits, where the gap to an all-digital MXFP4 baseline is $\leq 1\%$ across 3-5 CM-correction bits (Fig. \ref{fig:adc_accuracy_vitb}). Smaller ADCs (8 or 9 bits) incur large accuracy penalties (up to 13\%) in comparison.
We therefore adopt a \textbf{10-bit} SAR ADC in the CTT-CIM macros to support PTQ-only, drop-in deployment with near-digital accuracy.

There is a large area ($1.2\times$) and energy overhead ($2.9\times$) for a 10-bit ADC at the macro level (for a $1024\times1024$ macro, compared to a scaled 8-bit ADC \cite{adc_source}). To combat this, as mentioned in Section \ref{analog_stage}, we amortize ADC power and penalty by creating macros that are as tall as possible.

\section{MXFormer Architecture \& Dataflow}
\begin{figure*}[h!]
    \centering
\includegraphics[width=0.97\textwidth]{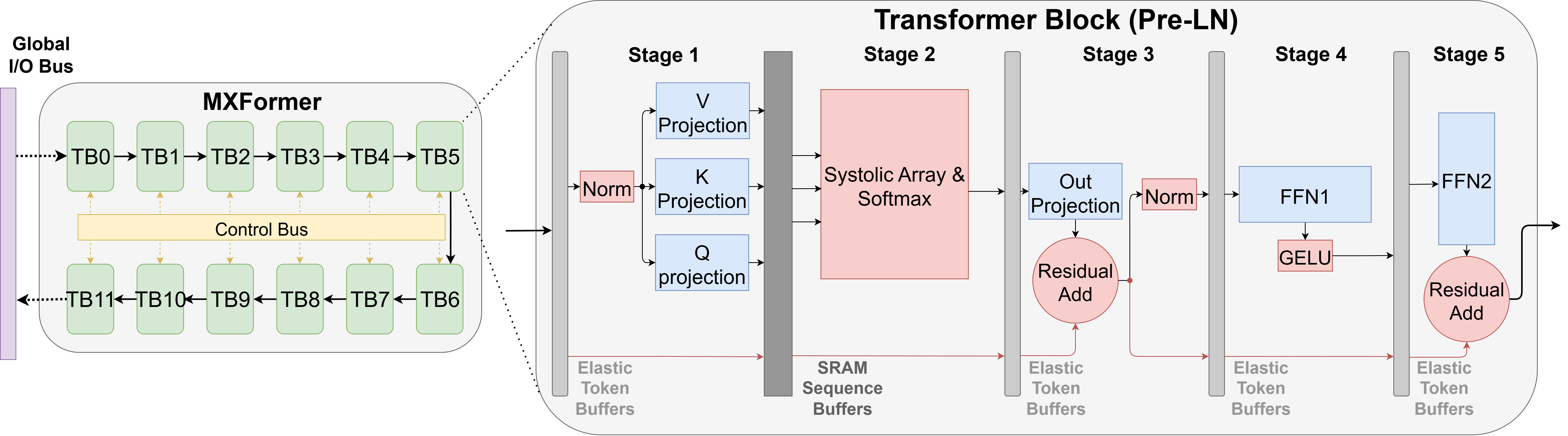}
\caption{The high-level system architecture. Within the 5-stage pipelined Transformer Block, blue rectangles represent CTT CIM arrays. Red rectangles represent digital elements.}
    \label{fig:system_architecture}
\end{figure*}

\subsection{MXFormer Architecture Overview}
Our accelerator is designed with a singular focus on maximizing single-stream throughput for fixed-model inference. The architecture's core philosophy is a deeply pipelined, fully weight-stationary dataflow. As shown in Fig.~\ref{fig:system_architecture}, the entire accelerator is organized as 12 identical Transformer Blocks, each of which processes one Transformer encoder/decoder layer in a pipelined fashion. An input sequence stream enters the first block, and the output of each block is passed directly to the input of the next, forming a macro-pipeline that spans the entire chip.

If the target model does not fit on a single chip, we statically shard its layers across multiple chips and pipeline them end-to-end, inserting the inter-chip link as an additional ``pipeline stage''.

\subsection{Pipeline Overview}
This pipeline design is physically realized through a static hardware partitioning strategy within each Transformer Block. The full list of stages and operations per stage is listed below ($^{\mathrm{A}}$ = analog (CTT), $^{\mathrm{D}}$ = digital):

\begin{enumerate}[leftmargin=*]
\item \textbf{Stage 1: Pre-Attention Prep}\\
\textbf{Ops:} LayerNorm$^{\mathrm{D}}$; linear projections for $\mathbf{Q},\mathbf{K},\mathbf{V}$ $^{\mathrm{A}}$

\item \textbf{Stage 2: Attention Core}\\
\textbf{Ops:} $\mathbf{S}=\mathbf{Q}\mathbf{K}^{\top}$ $^{\mathrm{D}}$; $\mathbf{O}=\mathrm{softmax}(\mathbf{S})\mathbf{V}$ $^{\mathrm{D}}$

\item \textbf{Stage 3: Attention Epilogue}\\
\textbf{Ops:} Output linear projection$^{\mathrm{A}}$; residual add$^{\mathrm{D}}$; LayerNorm$^{\mathrm{D}}$

\item \textbf{Stage 4: FFN Part 1}\\
\textbf{Ops:} First FFN$^{\mathrm{A}}$; GELU$^{\mathrm{D}}$

\item \textbf{Stage 5: FFN Part 2}\\
\textbf{Ops:} Second FFN$^{\mathrm{A}}$; residual add$^{\mathrm{D}}$
\end{enumerate}

For inter-stage buffering and flow control, there are two different levels of granularity: 
\begin{enumerate}
    \item \emph{Sequence-level}: the S1$\rightarrow$S2 handoff ($\mathbf{Q}$/$\mathbf{K}$ in MXFP4, $\mathbf{V}$ in INT10, and the input residual) is stored in a full-sequence double-buffered SRAM block so Stage 1 can write to the buffer while Stage 2 consumes the previous sequence. A full-sequence buffer is needed because of the Attention operation. Overall system throughput is set by the sequence-level latency.
    \item \emph{Token-level (streaming):} all other handoffs, including the inter-block/inter-chip S6$\rightarrow$S1  stream per-token through small two-entry elastic buffers to minimize storage area and power.
\end{enumerate}

\subsection{Analog Stages}
\label{analog_stage}
Each Transformer Block consists of 12 analog CTT CIM arrays. Four of these are used for the linear projections ($W_Q, W_K, W_V, W_O$), while the remaining eight are organized into two `large' arrays to implement the Feed-Forward Network (FFN) layers. Each large FFN array is composed of four smaller CTT arrays working in concert. For an optimal mapping of these static operations onto the analog arrays, each CTT array is sized to be $hidden\_dim$ $\times$ $hidden\_dim$. This minimizes the need for partial tile buffers and optimizes analog performance (taller arrays amortize fixed energy/area costs). The vector and ancillary units for all stages are explicitly provisioned so they never throttle the CTT arrays.

\subsection{Digital Stage}
For the digital stage, we selected conventional systolic arrays to guarantee fully accurate digital computation for all MXFP4 operations within the attention mechanism. This is a critical design choice, as digital CIM designs like SRAM CIM are noisy enough to necessitate Quantization-Aware Training (QAT) \cite{imc_qat_noise_1, imc_qat_noise_2}, which goes against our goals of a drop-in, no-retraining system.

\begin{figure}[!htbp]
\centering
\includegraphics[width=0.47\textwidth]{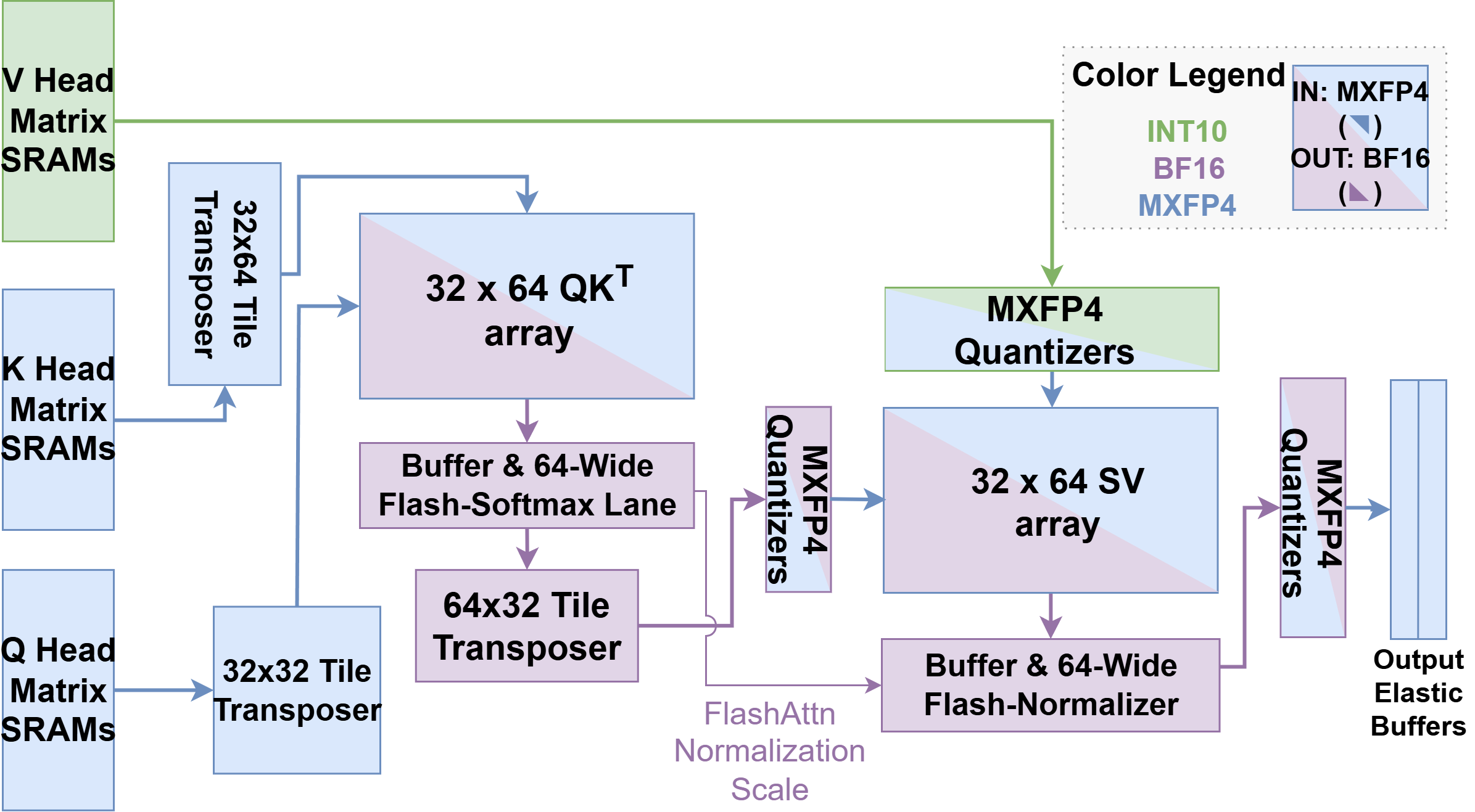}
\caption{High level view of the fully pipelined digital datapath in Stage 2. All blocks color coded according to the legend.}
\label{fig:attn-arch}
\end{figure}

A high-level overview of Stage 2 is visible in Fig. ~\ref{fig:attn-arch}. As shown, Stage 2 instantiates two \emph{output-stationary} ($32\times64$) systolic arrays: one for $QK^{\mathsf T}$ and one for $SV$. 

\begin{figure}[!htbp]
\centering
\includegraphics[width=0.42\textwidth]{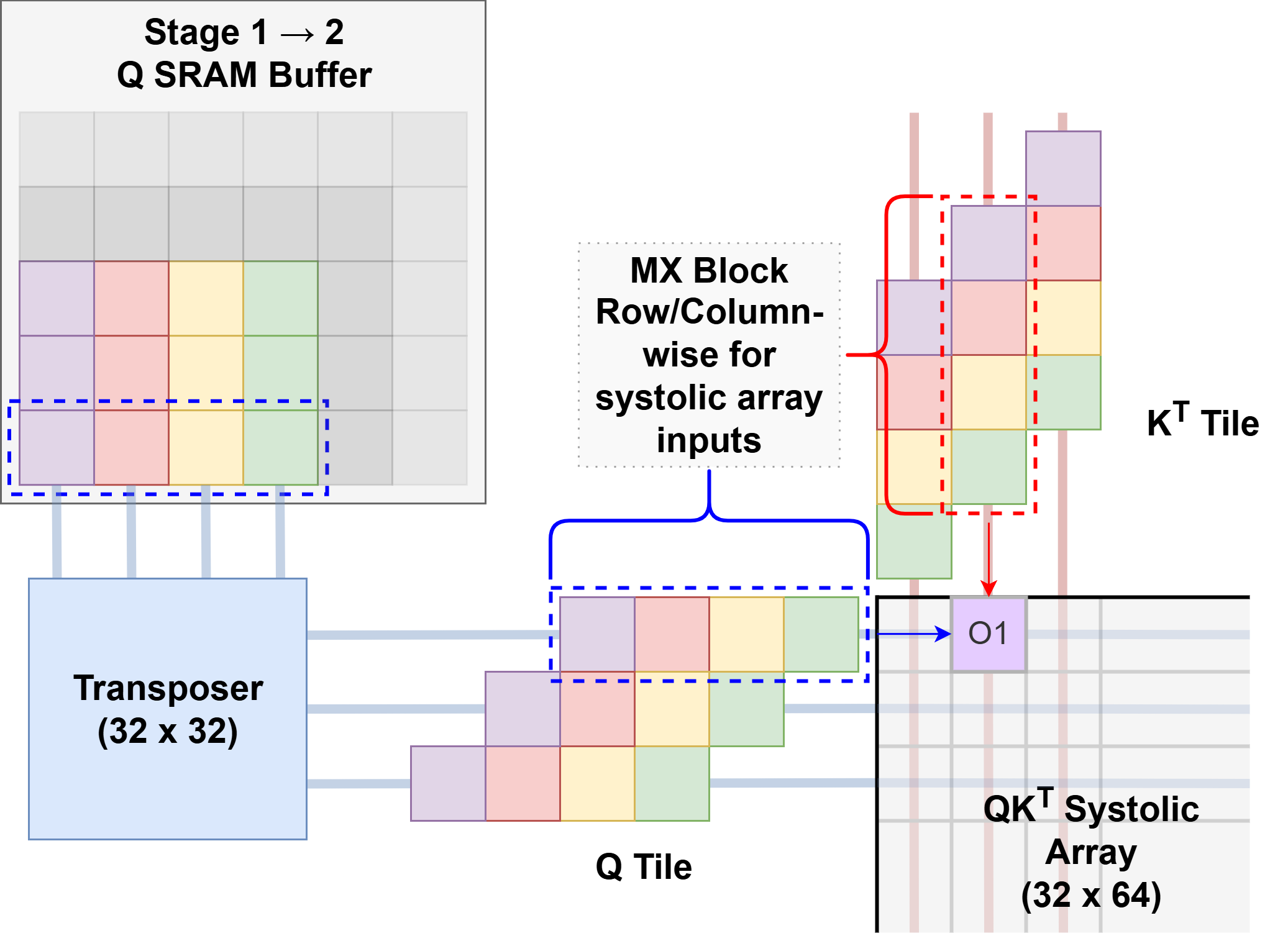}
\caption{In this illustration, element O1 in the systolic array ``sees'' only 2 MXFP blocks per 32 inputs/weights (the dotted blue [$\mathbf{Q}$] and red [$\mathbf{K^\mathsf T}$] MXFP blocks). $\mathbf{Q}$ blocks are quantized in a row-major direction (from the array's perspective), while  $\mathbf{K^\mathsf T}$ blocks are quantized in a column-major direction.
To enable this, the $\mathbf{Q}$ and $\mathbf{K}$ matrix have been MXFP-quantized in a row-major format.}
\label{fig:transpose-mxfp}
\end{figure}

As shown in Fig. ~\ref{fig:transpose-mxfp}, to ensure that every tile arriving at an array is both MXFP-quantized with the correct orientation and transposed appropriately, we carefully orchestrate storage, quantization, and movement across the pipeline using Gemmini \cite{gemmini} transposers and buffers (not pictured). $\mathbf{Q}$ and $\mathbf{K}$ use a $32\times32$ / $32\times64$ tile transposer respectively. As mentioned previously, $\mathbf{V}$ is stored in INT10 because it feeds the top of the $SV$ array and must be MXFP block-quantized along the \textbf{column} dimension. We perform the quantization \emph{in-situ} on a per-tile basis as $\mathbf{V}$ streams into $SV$. 
We also quantize the Softmax results \emph{in-situ} after a transpose operation as they feed the $SV$ array.

We do not explicitly utilize ``replay'' buffers to support reuse, as the SRAM blocks are highly energy efficient and are sized to supply the needed bandwidth.
The Microscaling shared exponents are transmitted on separate buses, following their respective blocks in a sparse fashion (1 exponent per block of 32).

Between the two arrays is a 64-wide Softmax lane designed to operate as a fully pipelined stage with no internal hardware reuse. To support sequence lengths $>64$ and minimize buffer storage, we adopt an in-hardware FlashAttention-style \cite{flashattn} deferred Softmax. We maintain a running maximum and sum across tiles, we scale outgoing tiles, and the final division is postponed until after the $S\times V$ multiply in a dedicated normalizer block. The Softmax lane also performs $1/\sqrt{d_k}$ scaling.

We choose $32\times64$ systolic arrays to balance area and performance for our target workloads and to align with the common ViT head dimension of 64, which simplifies tiling. All blocks, including SRAM access, transposers, quantizers, both arrays, the pipelined Softmax, and the normalizer, are throughput-matched and deeply pipelined, yielding bubble-free steady-state execution.

\subsection{Systolic Array Design}
\label{sec:systolic_array}
\begin{figure}[!htbp]
\centering
\includegraphics[width=0.42\textwidth]{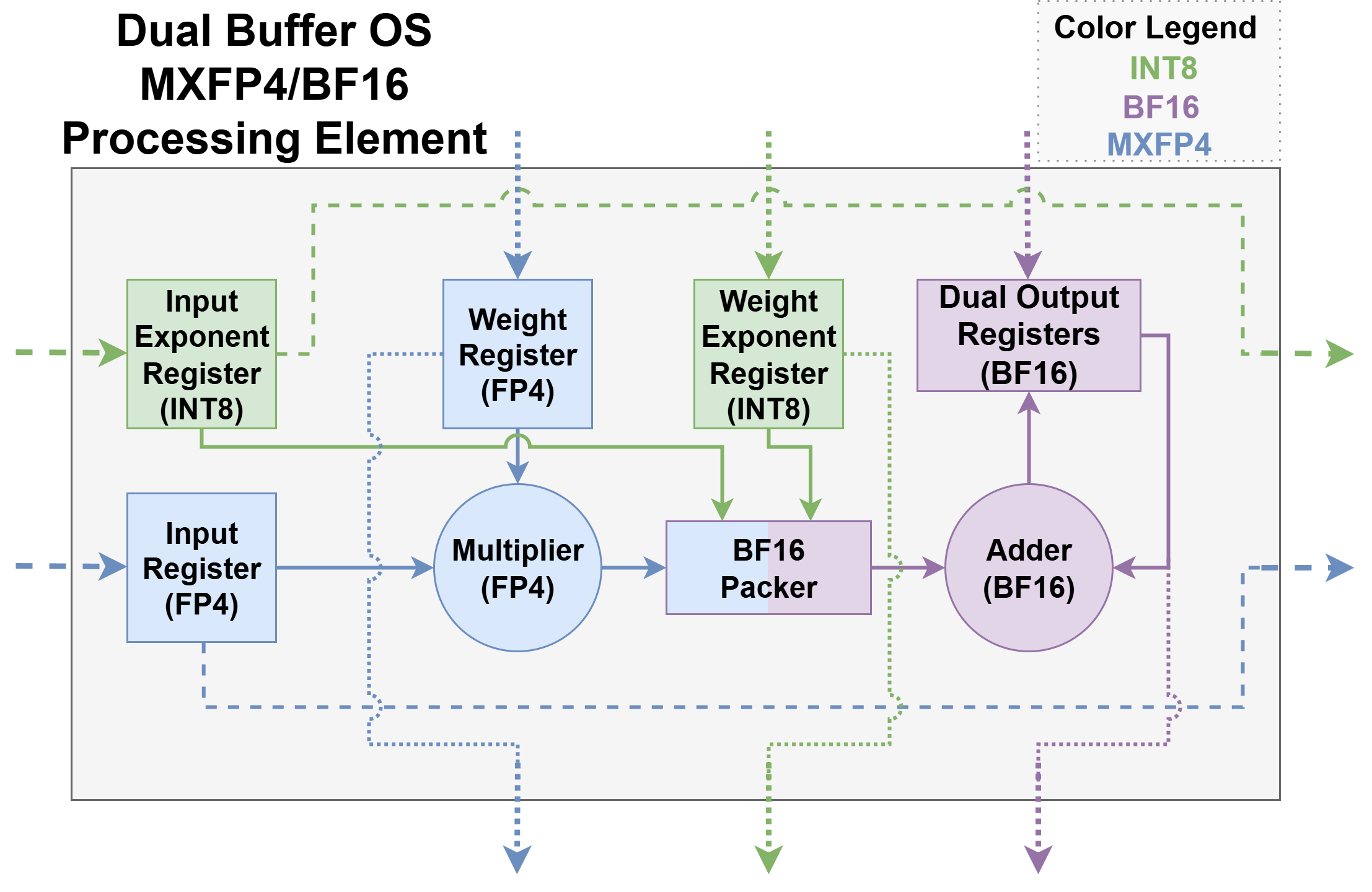}
\caption{Systolic Array MXFP-enabled PE. Behaves as a conventional, 100\% utilization \cite{gemmini} dual buffer PE. The BF16 packer block contains an internal INT8 adder (not shown). Blocks and wires color coded according to the legend.}
\label{fig:mxfp4_pe}
\end{figure}

The systolic arrays are heavily derived from Gemmini \cite{gemmini}. Our array processing element (PE), illustrated in Fig. \ref{fig:mxfp4_pe}, preserves the Gemmini FP4 multiplication and BF16 accumulation with dual-buffered output registers. We add a lightweight INT8 side path for \emph{MXFP4} shared exponent handling. Two small INT8 registers buffer the input and weight block exponents. Every multiply product is forwarded to a packer that forms a BF16 value by converting the product while also adding the sum of the input and weight exponents into the exponent field. The converted value is then accumulated in BF16 as usual. More details are in Appendix~\ref{app:conv}.

The \emph{in-situ} MXFP4 $\rightarrow$ BF16 conversion within the PE eliminates a key scheduling bottleneck. Deferring conversion until after the reduction prevents combining partial sums with different exponents. This would require the array to drain after each input/weight block, severely limiting reuse. Instead, \emph{in-situ} conversion allows the array to function like a dual-buffer systolic array with high utilization \cite{gemmini}. 
\section{MXFormer Evaluation and Results}
\label{sec:results}
All \emph{MXFormer} results assume a digital clock of $1\mathrm{GHz}$, an analog clock of $169\mathrm{MHz}$, a 10-bit ADC, a 3-bit CM correction budget, and the Row Hist 2-Pass strategy.
\subsection{Systems under test}
As shown in Table \ref{tab:systems-summary}, we evaluate two fixed accelerator configurations. Both configurations have a maximum sequence length of 512 due to limited buffer size.
As large models have 24 layers, they are mapped across \emph{two} identical \textbf{Large} dies, pipelined end-to-end. Only activations are streamed, so I/O overhead is minimal as quantified in Table~\ref{tab:all-models-metrics}.

\begin{table}[!htbp]
\centering
\small
\caption{Summary of systems under test. Metrics reported at peak TOPS (\textbf{Base}: seq len = 256, \textbf{Large}: seq len = 192).}
\label{tab:systems-summary}
{
\begin{tabular}{@{}lcccccc@{}}
\toprule
\textbf{System} & \textbf{\shortstack{Array\\Size}} & \textbf{\shortstack{Area\\(mm$^2$)}} & \textbf{\shortstack{Power\\(W)}} & \textbf{TOPS} & \textbf{\shortstack{TOPS\\/mm$^2$}} & \textbf{\shortstack{TOPS\\/W}} \\
\midrule
\textbf{Base}  & $768$  & 376.3 & 163.16 & 1515.14 & 4.04 & 9.29  \\
\textbf{Large} & $1024$ & 561.5 & 182.61 & 2631.56 & 4.69 & 14.41  \\
\bottomrule
\end{tabular}}
\end{table}

\subsection{Experimental Setup}
\subsubsection{Hardware Modeling}
\noindent \textbf{Digital blocks}: We implemented all digital components in RTL and synthesized them to the GF22FDX SOI standard-cell library to obtain Power-Performance-Area numbers. SRAM numbers were sourced from vendor-provided SRAM datasheets. Most of the components rely heavily on open-source RTL \cite{gemmini,flopoco_blocks,mx_to_fp_conv,hardfloat_repo,fpnew_blocks} (see Appendix \ref{app:rtl}). \\
\noindent \textbf{Analog blocks}: We modeled the analog blocks by extrapolating the post-layout results of a macro designed in GF22FDX SOI. Energy numbers are calculated from characterized and tested CTTs that were manufactured in GF22FDX SOI. Area numbers are obtained from the extrapolation of post-layout schematics. ADC numbers are extrapolated using area-scaling of a previously published SAR-ADC \cite{adc_source}. \\
\noindent \textbf{Systems}: The detailed area and power breakdown for the \textbf{Base} and \textbf{Large} systems can be seen in Table \ref{tab:area_power_breakdown_merged_compact}. Power is reported for peak TOPS, so identical components in the 2 systems have slightly different power consumptions. This phenomenon is caused by the complex interaction between digital power gating, system pipeline behavior, and systolic array tile quantization effects, which cause the digital arrays to be slightly underutilized at peak TOPS.

\begin{table}[!htbp]
\centering
\small
\caption{Area and power breakdown for \textbf{Base} and \textbf{Large} system. $^\ast$All vector units include internal MXFP quantizers. Power reported at peak TOPS for each system.}
\label{tab:area_power_breakdown_merged_compact}
\begingroup
\begin{tabular}{l r|r || r|r}
\toprule
 & \multicolumn{2}{c}{\textbf{Area (mm$^2$)}} & \multicolumn{2}{c}{\textbf{Power (W)}} \\
\cmidrule(lr){2-3}\cmidrule(lr){4-5}
\textbf{Component} & {\textbf{Base}} & {\textbf{Large}} & {\textbf{Base}} & {\textbf{Large}}\\
\hline
\textbf{Systolic Arrays} & 58.25 & 58.25 & 87.51 & 85.23 \\
\hspace{1em}Gemmini Array & 55.99 & 55.99 & 85.61 & 83.38 \\
\hspace{1em}MXFP additions & 2.26 & 2.26 & 1.90 & 1.85 \\
\textbf{Vector Compute$^\ast$} & 14.54 & 17.35 & 16.82 & 19.14 \\
\hspace{1em}Softmax & 6.22 & 6.22 & 9.16 & 8.92 \\
\hspace{1em}Adders & 1.10 & 1.42 & 0.88 & 1.18 \\
\hspace{1em}LayerNorm & 5.35 & 7.21 & 5.40 & 7.21 \\
\hspace{1em}GELU & 1.86 & 2.51 & 1.37 & 1.83 \\
\textbf{CTT Macros} & 256.30 & 427.70 & 48.93 & 67.80 \\
\textbf{MXFP Quantizers} & 7.89 & 7.89 & 6.99 & 6.91 \\
\textbf{Transposers} & 1.15 & 1.15 & 1.10 & 1.07 \\
\textbf{Buffers} & 2.05 & 2.73 & 1.70 & 2.26 \\
\textbf{SRAMs} & 34.98 & 46.43 & 0.12 & 0.20 \\
\hline
\textbf{Total} & \textbf{375.20} & \textbf{561.50} & \textbf{163.16} & \textbf{182.61} \\
\bottomrule
\end{tabular}
\endgroup
\end{table}

\subsubsection{Software Modeling}
\label{sec:software}
We used Microsoft’s Microscaling library \cite{microsoft_microscaling_lib} to obtain accurate MXFP4 software baselines. Because native MXFP4 checkpoints are scarce, we perform QAT \textbf{solely} to create MXFP4 reference models (not to tune for our hardware): (i) ViT-B/16 \cite{original_vit_paper} on Food-101 \cite{food101} (weights from \cite{vit_b16_food101_repo}), (ii) ViT-S/16 on ImageNet-1k \cite{imagenet} (weights from \cite{vit_s16_imagenet1k_repo}), and (iii) BERT-Base \cite{bert} on SQuAD v2 \cite{squad2} (weights from \cite{bert_base_cased_squad2_repo}).

We then extended the Microscaling library with an analog front-end model that mirrors the CTT-CIM datapath:
(i) per-block partial sums are aligned using the programmed target exponent and a limited current mirror shift budget (\emph{CM correction bits}). Saturation induces explicit underflow/overflow events at the block outputs.
(ii) 2-pass extends the range of the current mirror shift budget by performing 2 passes through the layer.
(iii) a lossy $n$-bit ADC quantizes column sums to produce final column outputs.
Results in Fig.~\ref{fig:adc_accuracy_vitb}, Fig.~ \ref{fig:strategy_comparison} and Table~\ref{tab:acc_res} use this framework.

\begin{table}[!htbp]
\centering
\footnotesize
\caption{Final model accuracy with \textbf{ADC = 10 bits} and \textbf{CM Correction bits = 3} using Row-Hist-2-Pass. MXFP4 accuracy was obtained from QAT training for MXFP4 (\textbf{only}). MXFormer result uses unaltered MXFP4 weights.}
\label{tab:acc_res}
\begin{tabular}{@{}l l c c@{}}
\toprule
\textbf{Model} & \textbf{Workload\cite{imagenet,squad2,food101}} & \textbf{MXFP4 (\%)} & \textbf{MXFormer (\%) } \\
\midrule
BERT\text{-}B      & SQuAD v2 (F1)         & 73.1 & 72.2  \\
ViT\text{-}S/16 (224px)    & ImageNet\text{-}1k (top-1)   & 73.2 & 72.8 \\
ViT\text{-}B/16 (224px)   & Food\text{-}101 (top-1)     & 82.9 & 82.4\\
\bottomrule
\end{tabular}
\end{table}

\subsubsection{System Modeling}
We model end-to-end runtime, area, and power consumption with a custom analytical Python flow. For the attention systolic arrays, it calls \textbf{ScaleSim} \cite{scalesim} to obtain tile-level timings. Because Scalesim does not natively model dual-buffer, fully pipelined arrays, we emulate steady-state overlap by inflating the GEMM workload.

\subsection{MXFormer system characterization}
As mentioned previously, linear ($O(N)$) layers map to \emph{analog} stages while attention ($O(N^2)$) maps to the \emph{digital} stage. As all analog stages take the same time to execute, mathematically, for a given fixed system, the steady-state period is $\mathcal{T}(N)=\max(c_{\mathrm{analog}} N, c_{\mathrm{digital}} N^2)$, so the throughput in ops/s is
\begin{equation}
\text{TOPS}(N)=\frac{\alpha N+\beta N^2}{\max(c_{\mathrm{analog}} N,c_{\mathrm{digital}} N^2)}
\end{equation}
TOPS peaks at the balanced midpoint where:
\begin{equation}
c_{\mathrm{analog}} N \approx c_{\mathrm{digital}} N^2 \longrightarrow
N_{\mathrm{balance}} \approx\ \frac{c_{\mathrm{analog}}}{c_{\mathrm{digital}}}
\end{equation}
Intuitively, the midpoint is where both analog and digital stages are fully utilized. For \textbf{Base}, $N_{\mathrm{balance}}\approx256$ and for \textbf{Large} $\approx192$ (matching the peak TOPS points in Table \ref{tab:systems-summary}).

\begin{figure}[!htbp]
    \centering
    \includegraphics[width=0.47\textwidth]{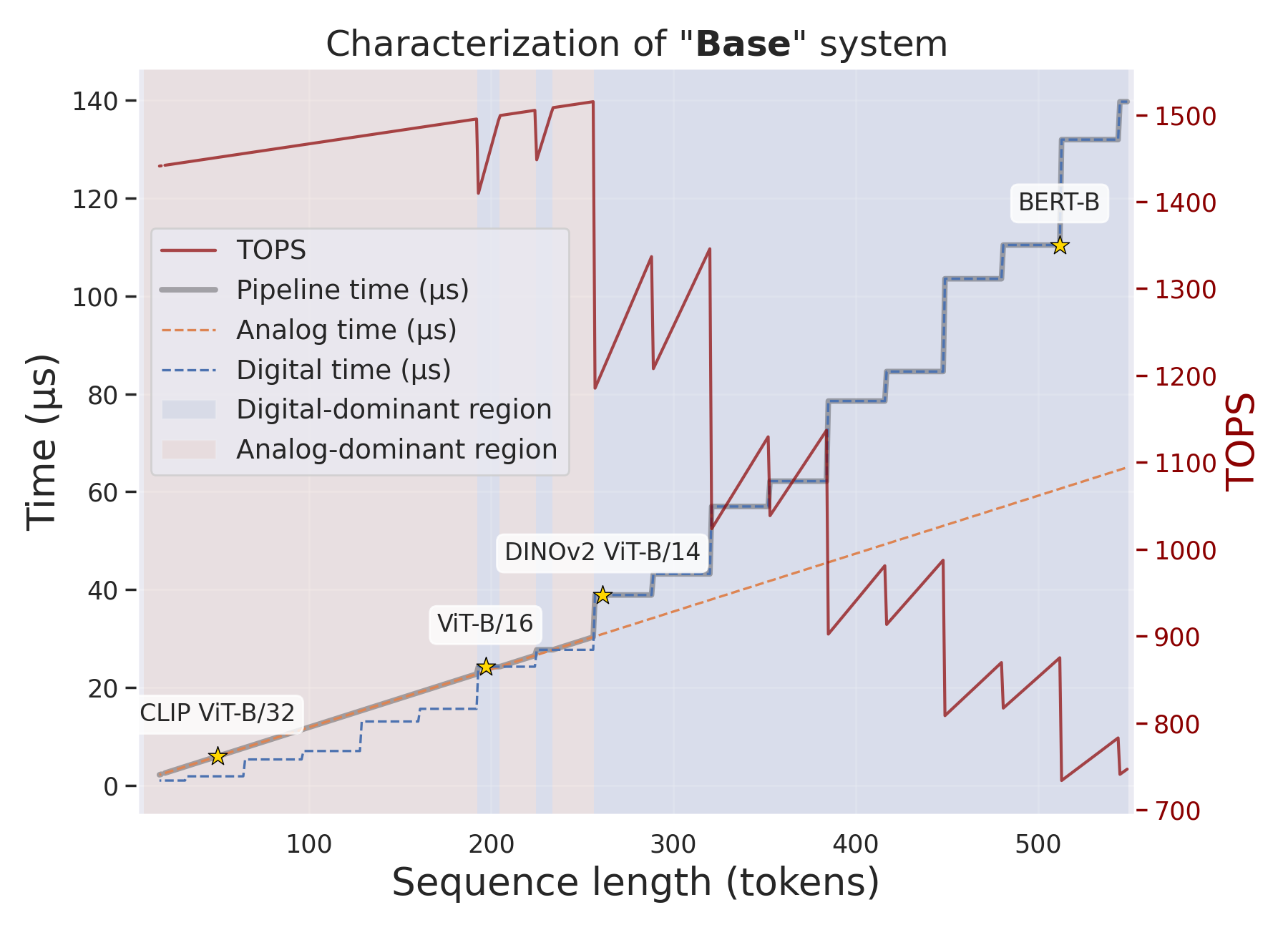}
    \caption{Base system characterization \textit{vs.} sequence length.}
    \label{fig:seq-sweep-system-tops}
\end{figure}

To showcase this effect, we now characterize the \textbf{Base} system as a function of sequence length. This applies to any model that fully utilizes the CTT arrays ($hidden\_dim = 768$). Fig. \ref{fig:seq-sweep-system-tops} shows the following:
(i) analog time, digital time, and the resultant pipeline time (left axis) and
(ii) sustained TOPS (right axis). Star markers denote real models evaluated at their canonical maximum sequence length.

In practice, because the systolic array geometry ($32{\times}64$ tiles) is comparable to sequence length, we observe significant distortive tiling quantization effects. Also, because of the static \textit{vs.} dynamic FLOPS imbalance (Fig. \ref{fig:static_vs_dynamic}),$
\frac{a}{c_{\mathrm{analog}}} \gg \frac{b}{c_{\mathrm{digital}}}$, which makes the TOPS drop-off dramatic past $N_{\mathrm{balance}}$. However, the overall mathematical trend still holds.

\subsection{Results}
Table~\ref{tab:all-models-metrics} summarizes per-model/per-system results at max sequence length. TOPS peaks near each system’s balance point (e.g., \textbf{ViT-L/32} on \textbf{Large}: 2612.0 TOPS), while very short-$N$ vision models emphasize FPS (e.g., \textbf{CLIP ViT-B/32} on \textbf{Base}: 169k FPS). Encoder NLP workloads (BERT-Base/Large), though attention-bound and outside our target regime, still reach $\geq$2.1 TOPS/mm$^2$ and $\geq$5.9 TOPS/W. Higher power than at peak TOPS is possible due to systolic array tile quantization artifacts.

Thanks to the deeply pipelined, fully weight-stationary design, \emph{MXFormer} always runs at the peak throughput for a given sequence length (Fig. \ref{fig:seq-sweep-system-tops}) and does \textit{not} suffer from scheduling/tiling or batch size bottlenecks that limit modern GPUs \cite{fastervit_a100_mfu_9_percent} or non-FWS accelerators. The only external requirement is modest activation I/O bandwidth, within even PCIe~3.0~$\times$16 ($\approx$16\,GB/s~\cite{pcie}), for all models in Table~\ref{tab:all-models-metrics}. We do not explicitly model external activation memory, but under a LPDDR4 assumption (19 pJ/bit\cite{amd_19pjbit_lpddr4}) the I/O power overhead is negligible ($<1\%$).

\subsection{Cross-Work and GPU Comparisons}

As shown in Tables~\ref{tab:all-models-metrics} and \ref{tab:comp_mixed_vert_lines_shrunk}, \emph{MXFormer} sustains \textbf{2.3-3.9}/ \textbf{2.1-4.7} TOPS/mm$^2$ and \textbf{3.1-14.5}/\textbf{7.8-13.5} TOPS/W (Base / Large), exceeding prior non-FWS photonic and digital/hybrid accelerators on both metrics. This demonstrates the benefits of a fully pipelined FWS architecture. Compared to IBM’s FWS 2D Mesh Analog Accelerator \cite{IBM_nvm_talk,ibm_pcm_pws} (14nm, PCM-based), \emph{MXFormer} (22nm) offers lower TOPS/W, but \textbf{$\sim$20.9$\times$} higher TOPS/mm$^2$, a gap driven by the density and low read latency of CTTs. Furthermore, it has \textbf{2}$\times$ the weight storage density, and does not require retraining for accurate model deployment. Lastly, due to \emph{MXFormer}'s high throughput, voltage scaling to improve efficiency is possible (we do not explore this in this work).

Table \ref{tab:mxformer_blackwell_compact} compares against an NVIDIA B200-class Blackwell GPU (5\,nm)\cite{nvidia_b200_hgx}. \emph{MXFormer} is markedly more efficient (\textbf{14.5} \textit{vs.}\ 9/4.5 TOPS/W) and can exceed compute density (\textbf{4.7} \textit{vs.}\ 1.13 TOPS/mm$^2$) in practice, despite using a 22nm technology node and being significantly smaller (per Table~\ref{tab:systems-summary}, \emph{MXFormer} TOPS/W and TOPS/mm$^2$ scale with area and array size).

\begin{table}[!htbp]
\centering
\footnotesize
\caption{Model results at max sequence lengths. For \textbf{Large}, models require 2 statically weight-sharded connected dies, so all metrics are for the combined system (I/O BW includes inter-chip comm.). \textsuperscript{$\dagger$} CLIP (vision path only) \cite{openaiCLIP}. \textsuperscript{$\ddagger$} DINOv2\cite{dinov2}. Input sizes: all ViTs use 224px except \textsuperscript{$\S$} which uses 384px. \textsuperscript{$\ast$} ViT-S/16 underutilizes the \textbf{Base} CTT arrays.}
\label{tab:all-models-metrics}
\begin{tabular}{lrrrrrr}
\toprule
\textbf{Model} & \shortstack{Power\\(W)} & \shortstack{Through-\\put} & \shortstack{TOPS} & \shortstack{TOPS\\/W} & \shortstack{TOPS\\/mm$^2$} & \shortstack{I/O\\BW\\(GiB/s)} \\
\midrule
\multicolumn{7}{l}{\textbf{Base}} \\
ViT\text{-}B/32\textsuperscript{$\dagger$}                 & 96.5  & 169000 & 1451 & 14.5 & 3.9 & 6.4 \\
ViT\text{-}B/16                                           & 170.6 &  41269 & 1440 &  8.4 & 3.8 & 6.2 \\
ViT\text{-}B/14\textsuperscript{$\ddagger$}               & 161.1 &  25716 & 1204 &  7.5 & 3.2 & 5.1 \\
BERT\text{-}Base                                          & 147.1 &   9055 &  875 &  5.9 & 2.3 & 3.5 \\
\textit{ViT-S/16}\textsuperscript{$\ast$}                 & 122.2 &  42893 &  389 &  3.1 & 1.0 & 3.2 \\
\midrule
\multicolumn{7}{l}{\textbf{Large} \textit{(2 chips)}} \\
ViT\text{-}L/32\textsuperscript{$\S$}                     & 385.5 &  58275 & 5224 & 13.5 & 4.7 & 12.8 \\
ViT\text{-}L/14\textsuperscript{$\dagger$}                & 327.4 &  19839 & 3208 &  9.8 & 2.9 & 7.7 \\
BERT\text{-}Large                                         & 299.2 &   6983 & 2338 &  7.8 & 2.1 & 5.4 \\
\bottomrule
\end{tabular}
\end{table}

\begin{table}[!htbp]
\centering
\small
\caption{GPU comparison. ``Peak'' uses B200’s theoretical MXFP4 9\,PFLOPS and a 1{,}000\,W budget (1,200\,W w/ HBM)\cite{toms_hw_gpu}. ``(ViT)'' assumes optimistic 20\% realized throughput (A100 ViT $\approx9\%$ \cite{fastervit_a100_mfu_9_percent}) and a conservative 2$\times$ efficiency penalty when downscaling FLOPS \cite{gpu_inefficiency}. For fairness, we report logic-only and HBM-inclusive figures \cite{toms_hw_gpu, micron_hbm3e}.}
\label{tab:mxformer_blackwell_compact}
\begin{tabular}{lccc}
\toprule
\textbf{Metric} & \textbf{\shortstack{MXFormer \\ Large}} & \textbf{\shortstack{B200 \cite{nvidia_b200_hgx} \\ Peak \textit{(ViT)}}} & \textbf{\shortstack{B200 w/HBM \\ Peak \textit{(ViT)}}} \\
\midrule
Tech. Node                 & 22\,nm & 5\,nm & 5\,nm \\
Area (mm$^2$)            & 561.5  & 1600  & 2568 \\
Power (W)                & 182.6  & 1000  & 1200 \\
TOPS (MXFP4)               & 2631.6   & 9000 (\textit{1800}) & 9000 (\textit{1800}) \\
TOPS/W                   & 14.41   & 9.0 (\textit{4.5})   & 7.5 (\textit{3.8}) \\
TOPS/mm$^2$              & 4.69    & 5.63 (\textit{1.13}) & 3.50 (\textit{0.7}) \\
\bottomrule
\end{tabular}
\end{table}

\begin{table*}[t]
\centering
\small
\caption{Comparison with academic SOTA for \textbf{D}eiT-B/16\cite{deit} and \textbf{B}ERT-Large (sequence length 128). For T-REX we show a projection to ``20nm'' using Stillmaker-Baas scaling in \textbf{bold} \cite{nm_scale}. For fairness, we do not scale non-digital systems. $^\ast$T-REX SoC BERT-L throughput (independent from other T-REX numbers).}
\label{tab:comp_mixed_vert_lines_shrunk}
\begin{tabular}{lccccc}
\toprule
\textbf{Metric} &
\shortstack{MXFormer\\ (Sim)} &
\shortstack{IBM 2\text{-}D Mesh\\Analog Accel \cite{IBM_nvm_talk,ibm_pcm_pws}\\(Sim)} &
\shortstack{Lightening \cite{photonic_accel_transformer} \\LT\text{-}L\text{-}4\\ (Sim)} &
\shortstack{T\text{-}REX\ \cite{t_rex_rawr} \\(Measured)} &
\shortstack{UCSD Hybrid\\Attn SoC  \cite{ucsd_hybrid_attn} \\(Measured)} \\
\midrule
Scope & Hybrid FWS SoC & Hybrid FWS/PWS SoC & Photonic accelerator & Digital accelerator & Hybrid SoC \\
\midrule
Tech. node & 22\,nm & 14\,nm & 14/16\,nm & 14 (\textbf{20})\,nm & 65\,nm \\
\midrule
Precision & MXFP4/BF16 & INT10/FP16 & Photonic (4\text{-}bit) & INT8 & INT4/INT8 \\
\midrule
Models & \shortstack{\textbf{D}eiT\text{-}B/16\\\textbf{B}ERT\text{-}L (N=128)} & \textbf{B}ERT\text{-}L (N=128) & \textbf{D}eiT\text{-}B/16 & \textbf{B}ERT\text{-}L (N=128) & Attention Only \\
\midrule
\shortstack{System Area\\(mm$^2$)} & \shortstack{\textbf{D:} 376.3\\\textbf{B:} 1123} & \shortstack{\textbf{B:} 350 (PWS)\\\textbf{B:} 2256  (FWS)} & \shortstack{\textbf{D:} 112.82} & \shortstack{\textbf{B:} 10.15 \textbf{(12.37)}} & \shortstack{3.2} \\
\midrule
Throughput & \shortstack{\textbf{D:} 41,269 img/s\\\textbf{B:} 8,449,997 tok/s} & \shortstack{\textbf{B:} 133,320 tok/s (PWS)\\\textbf{B:} 804,992 tok/s (FWS)} & \shortstack{\textbf{D:} 7507 img/s} & \textbf{B:} 2105 tok/s$^\ast$ & - \\
\midrule
Power (W) & \shortstack{\textbf{D:} 170.9\\\textbf{B:} 300.6} & \shortstack{\textbf{B:} 2.45 (PWS)\\\textbf{B:} 13.94 (FWS)} & \shortstack{\textbf{D:} 38.2} & \shortstack{\textbf{B:} 0.152\ \textbf{(0.095)}} & \shortstack{0.456} \\
\midrule
TOPS/mm$^2$ & \shortstack{\textbf{D:} 3.9 \\ \textbf{B:} 4.6} & \shortstack{\textbf{B:} 0.23 (PWS)\\\textbf{B:} 0.22 (FWS)} & \shortstack{\textbf{D:} 1.17} & \shortstack{\textbf{B:} 0.21\ (\textbf{0.076})} & \shortstack{0.079} \\
\midrule
TOPS/W & \shortstack{\textbf{D:} 8.5\\\textbf{B:} 17.3} & \shortstack{\textbf{B:} 33.6 (PWS)\\\textbf{B:} 35.5 (FWS)} & \shortstack{\textbf{D:} 3.45 } & \shortstack{\textbf{B:} 13.97\ (\textbf{9.9})} & \shortstack{0.56} \\
\midrule
Requires QAT & \textbf{No} & Yes & Yes & Yes & Yes \\
\midrule
Notes & \shortstack{DeiT-B w/ \textbf{Base}\\BERT-L w/ \textbf{2$\times$ Large}} & \shortstack{FWS and PWS from \cite{IBM_nvm_talk}} & \shortstack{LT-L-8 throughput used \\(Fig. inconsistent)} & \shortstack{Max freq.} & \shortstack{Max freq.} \\
\bottomrule
\end{tabular}
\end{table*}

\section{Comparison to Prior Work}

\noindent\textbf{TSMC microscaling gain-cell macro (ISSCC’25)~\cite{tsmc_macro}.} Implements lossless Microscaling CIM down to MXINT8 with strong macro TOPS/W and TOPS/$\mathrm{mm}^2$ (no SoC results), but weight \emph{storage} density is low ($\approx 34\mathrm{kb/mm}^2$). Our $1024{\times}1024$ CTT arrays reach $\approx 1756\mathrm{kb/mm}^2$ ($50\times$ higher), enabling full on-die weight residence. Consequently, this macro in a SoC would require PWS with large added bandwidth/energy cost, whereas \emph{MXFormer} enables a whole FWS system.

\noindent\textbf{HyFlexPIM (ISCA’25)~\cite{hyflexpim}.} Also FWS with ReRAM and is promising, but: (i) \emph{MXFormer} supports PTQ, whereas HyFlexPIM needs QAT and (ii) HyFlexPIM attention path uses digital ReRAM with only $\sim10^8$ write endurance, which, if used in \emph{MXFormer} with BERT-Base would be exhausted in $\approx 3.1$ hours under a very conservative assumption of one write per token-slot per sequence ($10^8/9052 \approx 184$ min.). \emph{MXFormer} uses conventional systolic arrays for attention (no endurance limits). A quantitative comparison isn’t possible since HyFlexPIM reports primarily relative metrics.

\noindent\textbf{Other FWS systems:} IBM’s PCM AIMC (PCM)~\cite{ibm_pcm_2023}, NeuRRAM (ReRAM)~\cite{neurrram_2022}, PUMA (ReRAM) \cite{puma_asplos19}, and ISAAC \cite{ISAAC} (Memristor). They demonstrate weight-resident execution but target CNNs or RNNs and remain in the $<\!100$M-parameter regime. None of them support Transformers at modern scales.

\section{Conclusion}
\emph{MXFormer} scales pipelined, fully weight-stationary MXFP4 execution to large ViTs in a reasonable area. \emph{MXFormer} \textbf{Large} holds a $307$M-parameter model \emph{on-die} across two 561.5mm$^2$ chips, forming an end-to-end cross-chip pipeline that sustains \textbf{58.3k FPS} on ViT-L/32. Relative to prior FWS designs, MXFormer reaches \textbf{20.9$\times$} higher TOPS/mm$^2$ and \textbf{2$\times$} higher storage density. Against comparable non-FWS accelerators, it delivers \textbf{$\approx$3.3$\times$--60.5$\times$} higher TOPS/mm$^2$ and \textbf{$\approx$1.7$\times$--2.5$\times$} better TOPS/W across representative models. All results are \textbf{PTQ-only} with $\leq1\%$ accuracy loss enabled by MXFP4-native CTT CIM with exponent alignment mechanisms and fully accurate digital attention compute.

\noindent \textbf{Future directions.} We target scaling along two axes: (i) greater resident capacity via higher CTT storage density, and (ii) more digital throughput to raise the optimally supported sequence length. On the device side, results indicate CTT scaling toward 14nm \cite{ctt_14nm}, improving density and efficiency. In parallel, a heterogeneous 3D-stacked architecture (advanced-technology-node digital compute bonded with 22nm/14nm analog CTT) offers a path to both larger on-die models and longer-sequence workloads. Target applications for a denser design may even include FWS execution of LLMs.

\pagebreak

\appendix
\section{Appendix: MXFP4$\leftrightarrow$BF16 Conversion Details}\label{app:conv}

The BF16 format, which uses the E8M7 encoding, can be cheaply and quickly converted to and from MXFP4, using the following operations: \\
\\
Let ($S_P$, $E_P$, $M_P$) be the bits of an E2M1 element. \\
Let ($S_{BF}$, $E_{BF}$, $M_{BF}$) be the bits of a BF16 number.\\
Let $E_X$ and $B_P$ be the bits of the E8M0 scale and the bias difference, respectively. \\
\\
$\textbf{MXFP4} \rightarrow$\textbf{BF16:} \quad
$\begin{cases} 
S_{BF} & \leftarrow S_P \\
E_{BF} & \leftarrow E_X + E_P - B_P \\
M_{BF} & \leftarrow M_P \ll 6
\end{cases}
\\[1em]
\textbf{BF16} \rightarrow \textbf{MXFP4:} \quad
\begin{cases} 
S_P & \leftarrow S_{BF} \\
E_X + E_P & \leftarrow E_{BF} + B_P \\
M_P & \leftarrow M_{BF} \gg 6 \quad (\text{truncation})
\end{cases}
$
\\
\\
\noindent For the systolic array PE \emph{in-situ} FP4$\times$FP4 multiplication, packing and conversion:
\\
$\textbf{MXFP4}\rightarrow \textbf{BF16:} \quad
\begin{cases}
S_{BF} \leftarrow S_{\text{product}} \\
E_{BF} \leftarrow (E_X{+}E_W) + (E_{x}{+}E_{w}) - 2B_P + \Delta \\
M_{BF} \leftarrow M_{\text{product}}
\end{cases}
$
\noindent Here $E_{x}$ and $E_{w}$ are the private Exponent bits of the activation and weight and $E_X$ and $E_W$ are their shared E8M0 block scales (our MXFP addition). 
$S_{\text{product}}$ and $M_{\text{product}}$ are the sign and 7-bit BF16 mantissa produced by the FP4 product and normalization operations. $\Delta$ is the small carry from that same normalization step.

\section{Appendix: RTL sources}\label{app:rtl}

Table \ref{tab:rtl-sources} shows the open source RTL libraries used for each digital component.

\newcommand{\rothead}[1]{\rotatebox[origin=c]{90}{\parbox{1.6cm}{\centering #1}}}

\begin{table}[!htbp]
\centering
\small
\caption{Digital components with external RTL sources. $^\ast$SoftmaxLane and LayerNorm: individual sub-blocks were synthesized only. Total PPA extrapolated with a conservative 20\% integration overhead. \,GELU: approximated as EXP LUT.}
\label{tab:rtl-sources}
\begin{tabular}{|l|c|c|c|c|c|}
\hline
\textbf{Component} &
\rothead{\textbf{Gemmini}\\\cite{gemmini}} &
\rothead{\textbf{FPNew}\\\cite{fpnew_blocks}} &
\rothead{\textbf{FloPoCo}\\\cite{flopoco_blocks}} &
\rothead{\textbf{HardFloat}\\\cite{hardfloat_repo}} &
\rothead{\textbf{mx-fpga}\\\cite{mx_to_fp_conv}} \\
\hline
Systolic Arrays & $\checkmark$ & & & $\checkmark$ & \\
\hline
Transposer & $\checkmark$ & & & & \\
\hline
Softmax Lane$^\ast$ & & $\checkmark$ & $\checkmark$ & & \\
\hline
LayerNorm$^\ast$ & & $\checkmark$ & $\checkmark$ & & \\
\hline
GELU$^\ast$ & & & $\checkmark$ & & \\
\hline
Res.\ Adder & & $\checkmark$ & & & \\
\hline
MXFP Quantizers & & & & $\checkmark$ & $\checkmark$ \\
\hline
\end{tabular}
\end{table}

\bibliographystyle{ACM-Reference-Format}
\bibliography{refs}

\end{document}